\documentclass{article}

\usepackage{PRIMEarxiv}

\usepackage{xcolor}
\usepackage{color}
\usepackage[utf8]{inputenc} %
\usepackage[T1]{fontenc}    %
\usepackage{hyperref}       %
\usepackage{url}            %
\usepackage{booktabs}       %
\usepackage{amsfonts}       %
\usepackage{nicefrac}       %
\usepackage{microtype}      %
\usepackage{lipsum}
\usepackage{fancyhdr}       %
\usepackage{graphicx}       %
\graphicspath{{media/}}     %

\usepackage{xcolor}
\usepackage{tcolorbox}
\usepackage{float}
\usepackage{amssymb}
\usepackage{multirow}
\usepackage[utf8]{inputenc}
\usepackage{tabularx}
\usepackage{booktabs} %
\usepackage{makecell}
\usepackage{subcaption}

\title{When LLMs Meet Cybersecurity: A Systematic Literature Review

}

\author{
  Jie Zhang$^{1,2}$, Haoyu Bu$^{1,2}$, Hui Wen$^{1,2,*}$, Yongji Liu$^{1,2}$, Haiqiang Fei$^{1,2}$, Rongrong Xi$^{1,2}$,\\
  \textbf{Lun Li$^{1}$, Yun Yang$^{1,2}$, Hongsong Zhu$^{1,2,*}$, Dan Meng$^{1,2}$} \\
  $^1$ Institute of Information Engineering, Chinese Academy of Sciences, Beijing, China \\
  $^2$ School of Cyber Security, University of Chinese Academy of Sciences, Beijing, China \\
  \texttt{\{zhangjie, wenhui, zhuhongsong\}@iie.ac.cn} \\
}

\begin{document}
\maketitle

\begin{abstract}

\noindent
The rapid development of large language models (LLMs) has opened new avenues across various fields, including cybersecurity, which faces an evolving threat landscape and demand for innovative technologies. Despite initial explorations into the application of LLMs in cybersecurity, there is a lack of a comprehensive overview of this research area. This paper addresses this gap by providing a systematic literature review, covering the analysis of over 300 works, encompassing 25 LLMs and more than 10 downstream scenarios. Our comprehensive overview addresses three key research questions: the construction of cybersecurity-oriented LLMs, the application of LLMs to various cybersecurity tasks, the challenges and further research in this area. This study aims to shed light on the extensive potential of LLMs in enhancing cybersecurity practices and serve as a valuable resource for applying LLMs in this field. We also maintain and regularly update a list of practical guides on LLMs for cybersecurity 
at \href{https://github.com/tmylla/Awesome-LLM4Cybersecurity}{https://github.com/tmylla/Awesome-LLM4Cybersecurity}.

\end{abstract}

\section{Introduction}
\label{sec:introduction}

Large language models (LLMs), represented by advanced models such as ChatGPT~\cite{ouyang2022training}, Llama~\cite{touvron2023llama}, and their derivatives~\cite{vicuna2023,almazrouei2023falcon,jiang2024mixtral} have marked a significant advancement in artificial intelligence. Leveraging massive data and advanced neural network architectures, these models have demonstrated remarkable capabilities in understanding and generating human language~\cite{zoph2022emergent,minaee2024large}. They not only set new benchmarks for achieving artificial general intelligence (AGI) but also show unique adaptability and effectiveness when collaborating with domain experts~\cite{ge2024openagi,kaur2024text}. Such research enables LLMs to be tailored to specific challenges in various fields, thereby promoting progress and development in areas such as healthcare, law, education, and software engineering~\cite{hou2023large,lai2024large,zhou2024survey,yan2024practical,li2023large,zhao2024chemdfm}. In the cybersecurity domain, exploring LLM applications can lay the foundations for further model development and utilization while highlighting potential transformative impacts~\cite{yao2024survey,das2024security,dasilva2024survey,motlagh2024large,yigit2024review}.

Cybersecurity is a critical issue given the growing number of cyber threats that pose significant risks to individuals, organizations, and governments~\cite{thakur2015investigation,scala2019risk,ghelani2022cyber}. The rapid evolution and dynamic nature of cybersecurity pose challenges as adversaries continuously adapt strategies to exploit vulnerabilities and evade detection~\cite{li2021comprehensive,aslan2023comprehensive}. While traditional approaches (\emph{e.g.}, signature-based detection, and rule-based systems) often struggle to keep pace with the evolving threat landscape, advancements in AI, particularly LLMs have opened new avenues for enhancing cybersecurity~\cite{ferrag2024generative}. 
On one hand, open-sourced LLMs (\emph{e.g.}, LLaMA~\cite{touvron2023llama,touvron2023llama2}) support the development of cybersecurity-enhanced domain LLMs such as RepairLlama~\cite{silva2024repairllama} and Hackmentor~\cite{hackmentor2023} to address unique cybersecurity challenges. On the other hand, advanced LLMs such as ChatGPT solve complex tasks via prompt engineering, in-context learning, and chains-of-thought despite the lack of cybersecurity-specific training~\cite{mohammed2024chatgpt}. These preliminary efforts show LLMs can aid cybersecurity tasks with promising results.

Despite the initial efforts of LLMs in cybersecurity, the field still faces several challenges~\cite{das2024security,pankajakshan2024mapping}. First, many studies rely on case studies without comprehensive methodology, raising concerns about scalability and reproducibility. In addition, the field lacks connectivity and in-depth analysis between studies. With the rapid increase in the amount of LLM research in this field, conducting a systematic overview is essential to guide the field into a new stage of development, in which the application of LLM is not just experimental but also has strategic impact~\cite{dasilva2024survey,motlagh2024large,yigit2024review}. Therefore, this work aims to conduct an extensive review of domain-specific LLMs tailored for cybersecurity, explore the breadth of LLM applications in this area, and identify emerging challenges to lay the foundation for future studies.

This survey aims to provide a comprehensive overview of the application of LLM in cybersecurity. We seek to address three key questions: 
\begin{itemize}
    \item RQ1: How to construct cybersecurity-oriented domain LLMs?
    \item RQ2: What are the potential applications of LLMs in cybersecurity?
    \item RQ3: What are the challenge and further research for the application of LLMs in cybersecurity?
\end{itemize}

\begin{table*}[t]
\centering
\vspace{-7pt}
\caption{\textbf{The main cybersecurity tasks and applications where LLMs have been utilized.}}
\vspace{1mm}
\label{table:content}
\setlength{\tabcolsep}{2.5mm}
\scalebox{0.85}{
    \begin{tabular}{cccccccccc}
    \toprule
        & 
        \textbf{\makecell[c]{Vulnerability\\Detection}} & 
        \textbf{\makecell[c]{(In)secure\\Code\\Generation}} & 
        \textbf{\makecell[c]{Program\\Repairing}} & 
        \textbf{\makecell[c]{Binary}} & 
        \textbf{\makecell[c]{IT\\Operations}} & 
        \textbf{\makecell[c]{Threat\\Intelligence}} & 
        \textbf{\makecell[c]{Anomaly\\Detection}} & 
        \textbf{\makecell[c]{LLM\\Assisted\\Attack}} & 
        \textbf{\makecell[c]{Others}} \\
    \midrule
        \textbf{RQ1} & $\checkmark$ & $\checkmark$ & $\checkmark$ & $\checkmark$ & $\checkmark$ & - & - & - & $\checkmark$ \\ 
        \textbf{RQ2} & $\checkmark$ & $\checkmark$ & $\checkmark$ & - & - & $\checkmark$ & $\checkmark$ & $\checkmark$ & $\checkmark$ \\ 
        \textbf{RQ3} & - & - & - & - & $\checkmark$ & - & $\checkmark$ & $\checkmark$ & - \\
    \bottomrule
    \end{tabular}
}
\end{table*}

By exploring these questions, we aim to bridge the gap between the advancement in LLMs and its potential impact on enhancing cybersecurity practices. We will delve into various cybersecurity tasks and applications to which LLMs are applicable, including vulnerability detection, secure code generation, program repair, binary, IT operations, threat intelligence, anomaly detection, and LLM-assisted attack, as shown in Table \ref{table:content}.

For the first question, we summarize the principles of existing cybersecurity LLMs, detailing their key techniques, the data used for model construction, and well-trained domain LLMs for special tasks. We provide insights into constructing domain models, which are valuable for researchers and practitioners looking to build customized LLMs based on specific requirements, such as computational limits, private data, and local knowledge bases (Section~\ref{sec:q1}). 
For the second question, we conduct an extensive survey on the usage of existing LLMs in more than 10 cybersecurity tasks, including threat intelligence, vulnerability detection, program repairing, and others. This analysis not only helps us understand how LLMs benefit cybersecurity in various aspects but also allows us to identify their strengths when applied to domain-specific tasks. By demonstrating the diverse capabilities of LLMs, we aim to illustrate their potential to enhance and transform the cybersecurity field (Section~\ref{sec:q2}).
The third question highlights the challenges that need to be overcome when applying LLMs in cybersecurity. LLMs' inherent vulnerabilities and susceptibilities lead to these attack challenges, especially attacks against LLMs and LLM jailbreaking. Additionally, we also explore further research directions for applying LLM to cybersecurity, guiding researchers and practitioners to promote advancement in this field (Section~\ref{sec:q3}).

In summary, this paper contributes by providing a comprehensive review of the state-of-the-art LLM applications in cybersecurity, highlighting the potential advantages and challenges, and proposing future research directions. The subsequent sections of this paper are organized as follows. 
Section \ref{sec:preliminary} outlines the scope of this paper. 
Section \ref{sec:q1} summarizes existing LLMs for cybersecurity. 
Section \ref{sec:q2} details how LLMs can be applied to various cybersecurity tasks. 
Section \ref{sec:q3} highlights the challenges and promising opportunities for future research. 
Section \ref{sec:conclusion} draws the conclusion.

\section{Preliminary}
\label{sec:preliminary}

\begin{figure}[htbp]
    \centering
    \begin{subfigure}[b]{0.45\textwidth}
        \includegraphics[width=\textwidth]{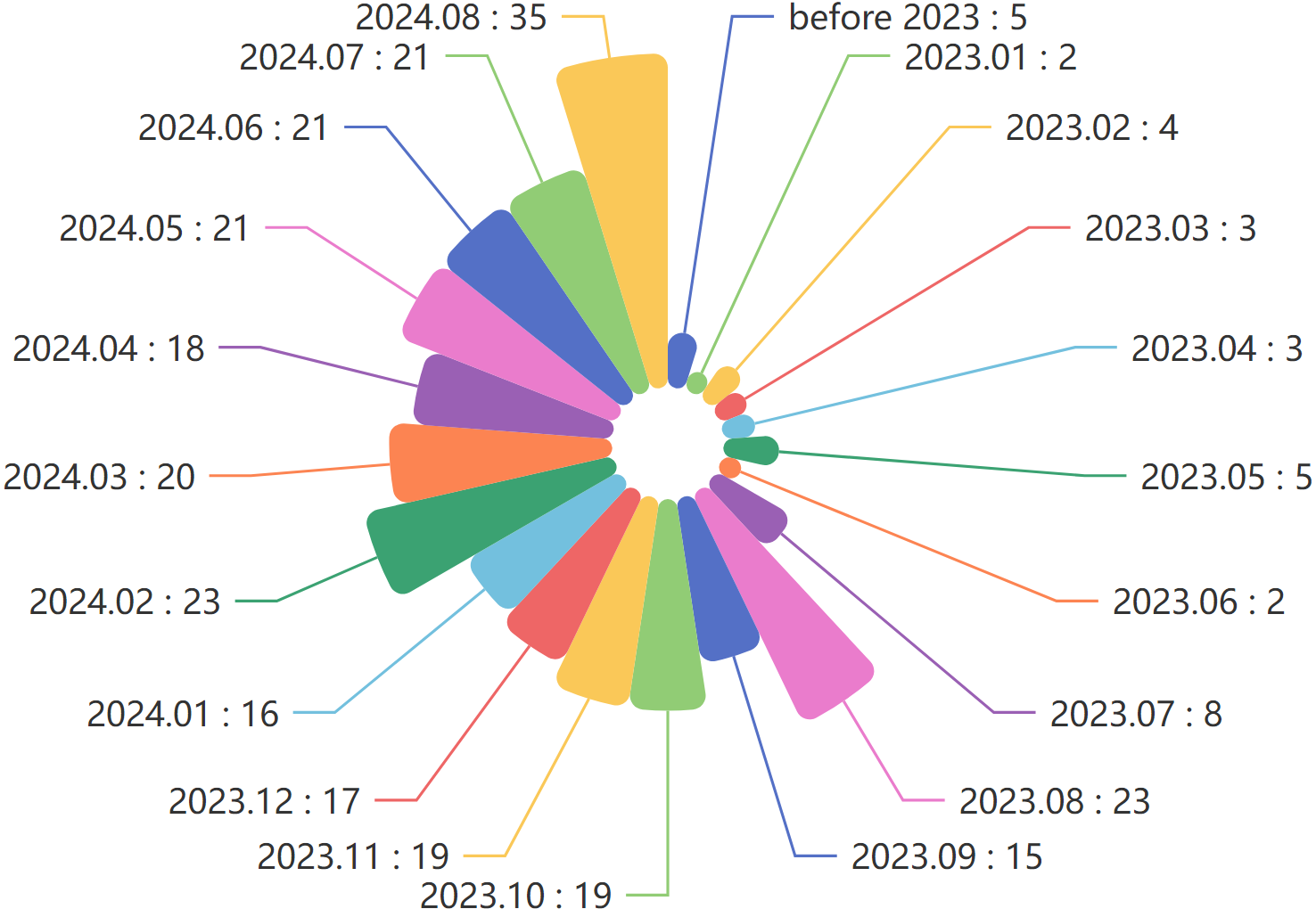}
        \caption{Time distribution over months}
        \label{fig:count}
    \end{subfigure}
    \hfill
    \begin{subfigure}[b]{0.45\textwidth}
        \includegraphics[width=\textwidth]{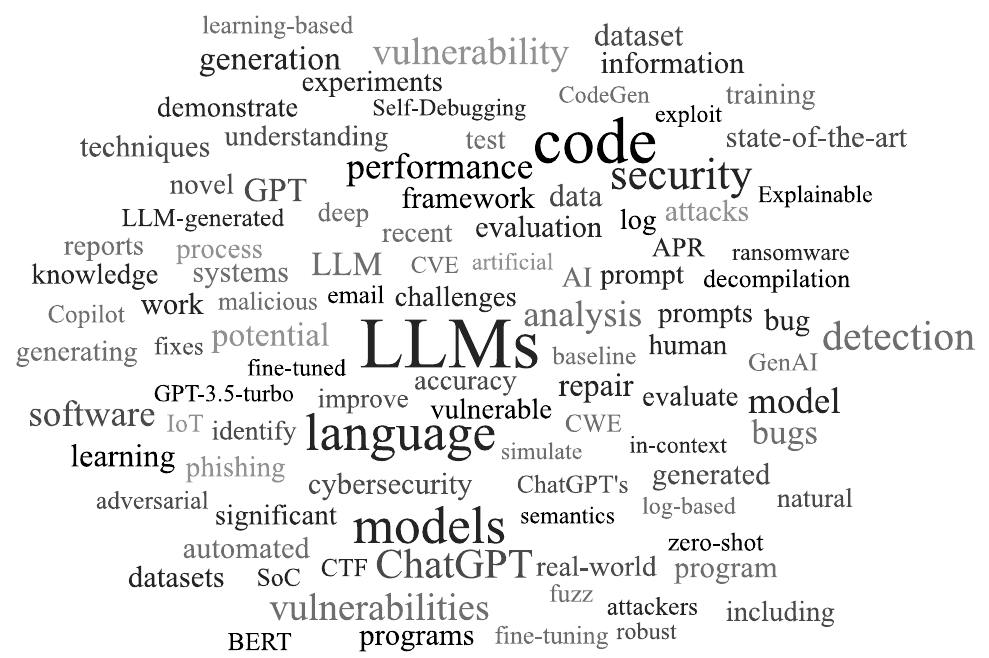}
        \caption{Word cloud}
        \label{fig:wordcloud}
    \end{subfigure}
    \caption{\textbf{Statistic of surveyed papers.}}
    \label{fig:statistic}
\end{figure}

In this review paper, we systematically investigate the application progress of LLMs in cybersecurity, covering more than 300 academic papers since 2023. Through an exhaustive study and comprehensive analysis, we aim to provide a detailed overview of the current state, challenges, and future directions of LLM applications in cybersecurity. As shown in Figure~\ref{fig:statistic}, this emerging research field continues to gain attention, and LLM can be used to solve various tasks. This not only highlights the current and potential impact of LLMs in cybersecurity, but also offers practical guidance for future research. Therefore, this section first summarizes the surveyed papers from two aspects: one is the LLMs used in cybersecurity, and the other is the category of cybersecurity tasks to which LLMs can be applied.

\subsection{LLMs in Cybersecurity}

LLMs have emerged as a transformative technology in the field of artificial intelligence, demonstrating remarkable capabilities in natural language understanding, generation, and reasoning~\cite{brown2020language,zoph2022emergent,minaee2024large}. These models, trained with large amounts of data, have the potential to revolutionize various fields, including the critical area of cybersecurity~\cite{motlagh2024large,yigit2024review}. The application of LLMs in cybersecurity is expected to enhance threat detection, automated vulnerability analysis, intelligent defense mechanisms, and more.

LLMs can be categorized into two main types: open-source and closed-source models. Open-source LLMs (\emph{e.g.}, Llama~\cite{touvron2023llama} and Mixtral~\cite{jiang2024mixtral}) provide model weights, and researchers can fine-tune the models for specific cybersecurity tasks. This adaptability is particularly valuable in cybersecurity scenarios, such as private data and models fine-tuned to customized needs. However, open-source LLMs may lack the performance and scale of closed-source LLMs.
On the other hand, closed-source LLMs (often referred to as commercial LLMs, \emph{e.g.}, ChatGPT~\cite{ouyang2022training} and Gemini~\cite{geminiteam2023gemini}), provide state-of-the-art performance and are maintained by commercial entities, often with access restrictions. While these models excel in accuracy and efficiency, their lack of transparency can raise concerns about potential biases and limitations in cybersecurity applications.

In the field of cybersecurity, there is a growing need for intelligent tools that can understand, analyze, and generate secure code. Code-based LLMs (\emph{e.g.}, CodeLlama~\cite{roziere2023code} and StarCoder~\cite{li2023starcoder,lozhkov2024starcoder}) are particularly well suited to address this demand. Unlike text-based LLMs are trained on vast amounts of natural language data, code-based LLMs are specifically designed to understand and work with programming languages. Code-based LLMs are trained on large code bases covering multiple programming languages, allowing them to capture the complexity of syntax, semantics, and common coding patterns. This specialized training enables them to perform a variety of tasks, including code completion, bug detection, and automated code review. In the context of cybersecurity, these capabilities are useful for identifying potential vulnerabilities, suggesting secure coding practices, and remediating security vulnerabilities.

\begin{table*}[htbp]
    \centering
    \caption{\textbf{A Summary of LLMs used in cybersecurity (this paper).}}
    \label{tab:llms}
    \vspace{1mm}
    \setlength{\tabcolsep}{2.5mm}
    \scalebox{0.85}{
        \begin{tabular}{@{}cccccc@{}}
        \toprule
        \multicolumn{1}{c}{\textbf{Organization}} &
          \multicolumn{1}{c}{\textbf{LLMs}} &
          \multicolumn{1}{c}{\textbf{Size}} &
          \multicolumn{1}{c}{\textbf{Open-Source}} &
          \multicolumn{1}{c}{\textbf{Count}} &
          \multicolumn{1}{c}{\textbf{Link}} \\ \midrule
        \multirow{4}{*}{OpenAI}      & GPT-3.5     & 175B             & $\times$      & 86  & \href{https://chat.openai.com/}{https://chat.openai.com/} \\
                                     & GPT-4       & -                & $\times$      & 56  & \href{https://chat.openai.com/}{https://chat.openai.com/} \\
                                     & Codex       & -                & $\times$      & 13   & \href{https://openai.com/blog/openai-codex}{https://openai.com/blog/openai-codex} \\
                                     & davinci(-002,-003)       & 175B   & $\times$      & 9   & \href{https://openai.com/blog/openai-api}{https://openai.com/blog/openai-api} \\
        \hline
        \multirow{2}{*}{Google}      & Bard\&Gemini        & -                & $\times$      & 12      & \href{https://gemini.google.com/}{https://gemini.google.com/} \\
                                     & PaLM(-1,-2)       & 540B                & $\times$      & 7      & \href{https://ai.google.dev/models/palm}{https://ai.google.dev/models/palm} \\
        \hline
        \multirow{1}{*}{Anthropic}   & Claude(-1,-2)    & -                & $\times$      & 2  & \href{https://claude.ai/}{https://claude.ai/} \\
        \hline
        \multirow{1}{*}{Github}      & Copilot    & -                & $\times$      & 2  & \href{https://github.com/features/copilot}{https://github.com/features/copilot} \\
        \hline
        \multirow{1}{*}{Microsoft}   & BingChat    & -                & $\times$      & 2  & \href{https://www.bing.com/chat}{https://www.bing.com/chat} \\
        \hline
        \multirow{2}{*}{EleutherAI}   & GPT-J    & 6B                & $\checkmark$      & 2  & \href{https://huggingface.co/EleutherAI/gpt-j-6b}{https://huggingface.co/EleutherAI/gpt-j-6b} \\
                                     & GPT-Neo    & 2.7B              & $\checkmark$      & 3  & \href{https://huggingface.co/EleutherAI/gpt-neo-2.7B}{https://huggingface.co/EleutherAI/gpt-neo-2.7B} \\
        \hline
        \multirow{3}{*}{Meta}       & Llama(-1,-2)       & 7B/13B/70B  & $\checkmark$  & 38  & \href{https://huggingface.co/meta-llama}{https://huggingface.co/meta-llama} \\
                                    & LlamaGuard    & 7B                & $\checkmark$      & 1      & \href{https://huggingface.co/meta-llama/LlamaGuard-7b}{https://huggingface.co/meta-llama/LlamaGuard-7b}\\
                                    & InCoder    & 1B/6B                & $\checkmark$      & 4      & \href{https://huggingface.co/facebook/incoder-1B}{https://huggingface.co/facebook/incoder-1B}\\
        \hline
        \multirow{1}{*}{LMSYS}       & Vicuna          & 7B/13B          & $\checkmark$      & 12  & \href{https://huggingface.co/lmsys/vicuna-7b-v1.5}{https://huggingface.co/lmsys/vicuna-7b-v1.5} \\
        \hline
        \multirow{1}{*}{LianjiaTech}  & BELLE     & 7B/13B    & $\checkmark$      & 1      & \href{https://github.com/LianjiaTech/BELLE/}{https://github.com/LianjiaTech/BELLE/}\\
        \hline
        \multirow{1}{*}{Databricks}   & Dolly     & 6B          & $\checkmark$      & 3      & \href{https://huggingface.co/databricks/dolly-v1-6b}{https://huggingface.co/databricks/dolly-v1-6b}\\
        \hline
        \multirow{1}{*}{-}  & Guanaco     & 7B                & $\checkmark$      & 2      & \href{https://huggingface.co/JosephusCheung/Guanaco}{https://huggingface.co/JosephusCheung/Guanaco}\\
        \hline
        \multirow{2}{*}{Salesforce}  & CodeGen(-1,-2)  & 3B/7B/16B    & $\checkmark$      & 9      & \href{https://github.com/salesforce/CodeGen/}{https://github.com/salesforce/CodeGen/}\\
                                    & CodeT5     & 6B                & $\checkmark$      & 3      & \href{https://huggingface.co/Salesforce/codet5p-6b}{https://huggingface.co/Salesforce/codet5p-6b}\\
        \hline
        \multirow{1}{*}{BigCode}     & StarCoder(-1,-2)     & 3B/7B/15B     & $\checkmark$     & 3      & \href{https://huggingface.co/bigcode/}{https://huggingface.co/bigcode/}\\
        \hline
        \multirow{1}{*}{THUDM}       & ChatGLM     & 6B                & $\checkmark$      & 8      & \href{https://github.com/THUDM/ChatGLM-6B}{https://github.com/THUDM/ChatGLM-6B}\\
        \hline
        \multirow{1}{*}{KaistAI}       & Prometheus     & 7B/13B       & $\checkmark$      & 1      & \href{https://github.com/kaistAI/Prometheus}{https://github.com/kaistAI/Prometheus}\\
        \hline
        \multirow{2}{*}{MistralAI}   & Mistral     & 7B                & $\checkmark$      & 6      & \href{https://huggingface.co/mistralai/Mistral-7B-v0.1}{https://huggingface.co/mistralai/Mistral-7B-v0.1}\\
                                     & Mixtral    & 8*7B              & $\checkmark$      & 5  & \href{https://huggingface.co/mistralai/Mixtral-8x7B-v0.1}{https://huggingface.co/mistralai/Mixtral-8x7B-v0.1} \\
        \hline
        \end{tabular}
    }
\end{table*}

\subsection{Cybersecurity Categories of LLMs Application}

\begin{figure}[htbp]
    \centering 
    \includegraphics[width=0.8\textwidth]{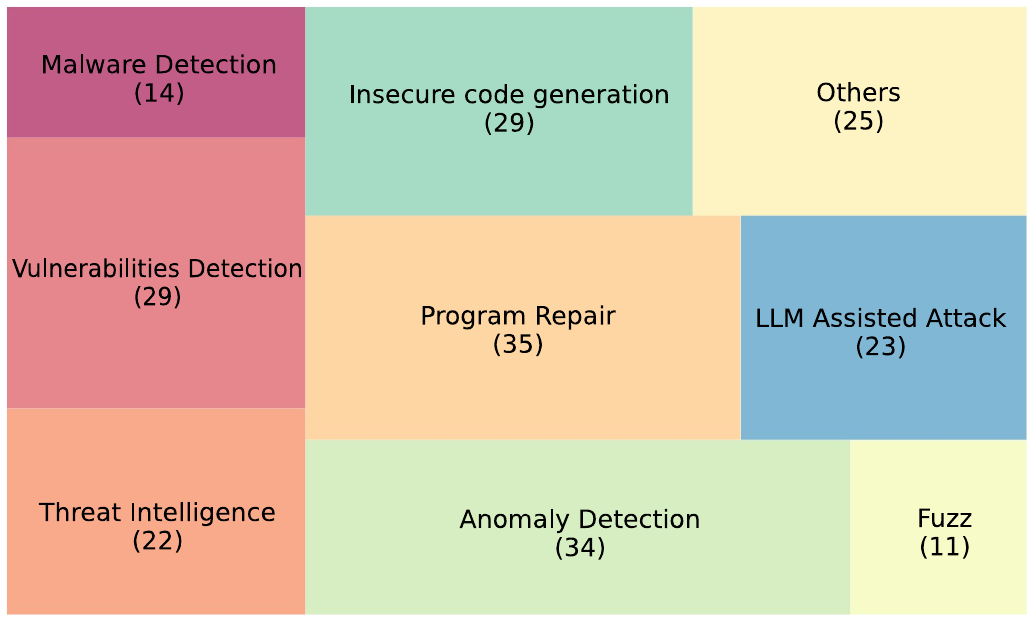}
    \caption{\textbf{Treemap for cybersecurity categories of LLMs' application.}}
    \label{fig:treemap}
\end{figure}

Cybersecurity has become a critical concern due to the increasing reliance on interconnected systems and the continued emergence of sophisticated cyber threats~\cite{thakur2015investigation,ghelani2022cyber}. The field of cybersecurity encompasses a wide range of practices, technologies, and strategies aimed at protecting computer systems, networks, and data from unauthorized access, attacks, damage, or disruption~\cite{li2021comprehensive,aslan2023comprehensive}. AI techniques, especially LLMs, have shown great potential in revolutionizing various aspects of cybersecurity~\cite{yigit2024review}. The applications of LLMs in cybersecurity are wide-ranging, including threat intelligence, vulnerability detection, malware detection, and anomaly detection, fuzz and program repair, LLM assisted attack(in)secure code generation, and others.

\begin{itemize}
    \item Threat Intelligence: It is very difficult to extract information from a large number of threat intelligence documents. Some researchers turn to LLMs to organize and analyze these massive and cluttered data.

    \item Vulnerability Detection: This is a critical task in cybersecurity, and has seen novel approaches emerge through the integration of LLMs.

    \item Malware Detection: LLMs can serve as both the static analysis assistant and the dynamic debugging assistant, improving the efficiency and effectiveness of the process.
    
    \item Anomaly Detection: It mainly refers to security anomalies such as malicious traffic in the flow, virus files in the system, anomalies in logs, etc.

    \item Fuzz: Traditional fuzzing techniques are effective in discovering software vulnerabilities, but their inherent limitations can affect their efficiency and effectiveness. The LLM-based approach for fuzzing is a promising area of research.
      
    \item Program Repair: Program repair is task-intensive and patching defects requires sufficient experience and knowledge. Many studies have proved the effectiveness of LLMs about this issue.

    \item LLM-Assisted Attacks: Many are not satisfied with LLMs' positive applications. They have discovered the effectiveness of LLMs in launching network attacks such as phishing emails and penetration testing.
    
    \item (In)secure Code Generation: Is there a risk in the code generated by LLMs? Moreover, can LLMs correct their code through some strategies?
    
    \item Others: In addition to the aspects mentioned above, we have also collected some researches which prove the importance of LLMs in the field of cybersecurity, there are fewer application studies of LLM in its field.
\end{itemize}

\section{RQ1: How to construct cybersecurity-oriented domain LLMs?}
\label{sec:q1}

The cybersecurity domain is facing escalating threats, demanding intelligent and efficient solutions to deal with complex and evolving attacks~\cite{kaur2023artificial,kumar2023artificial,mijwil2023towards}. LLMs provide new opportunities for the cybersecurity community~\cite{dasilva2024survey,motlagh2024large}. Trained on massive data, LLMs have acquired rich knowledge and developed strong understanding and reasoning capabilities, providing powerful decision-making for cybersecurity.

Advancing cybersecurity requires LLMs tailored to the field, leveraging their potential to learn domain-specific data and knowledge. This section firstly introduce several domain datasets for evaluating the cybersecurity capabilities of LLMs~\cite{tihanyi2024cybermetric,bhatt2023purple,tony2023llmseceval}, which can guide for selecting an appropriate LLM as the base model when constructing cybersecurity LLMs.
Then, we focus on key technologies for constructing cybersecurity LLMs, including training methods such as continual pre-training (CPT)~\cite{yıldız2024investigating,zhang2024leveraging} and supervised fine-tuning (SFT)~\cite{zhang2023instruction,dong2023abilities} of LLMs, as well as technical implementations like full-parameters fine-tuning and parameter-efficient fine-tuning (PEFT)~\cite{ding2023parameter}. Finally, we summarize existing customized LLMs for specific cybersecurity tasks~\cite{ferrag2023securefalcon,hackmentor2023}, including vulnerability detection, program repair, secure code generation, etc.

\begin{figure*}[htbp]
    \centering 
    \includegraphics[width=\textwidth]{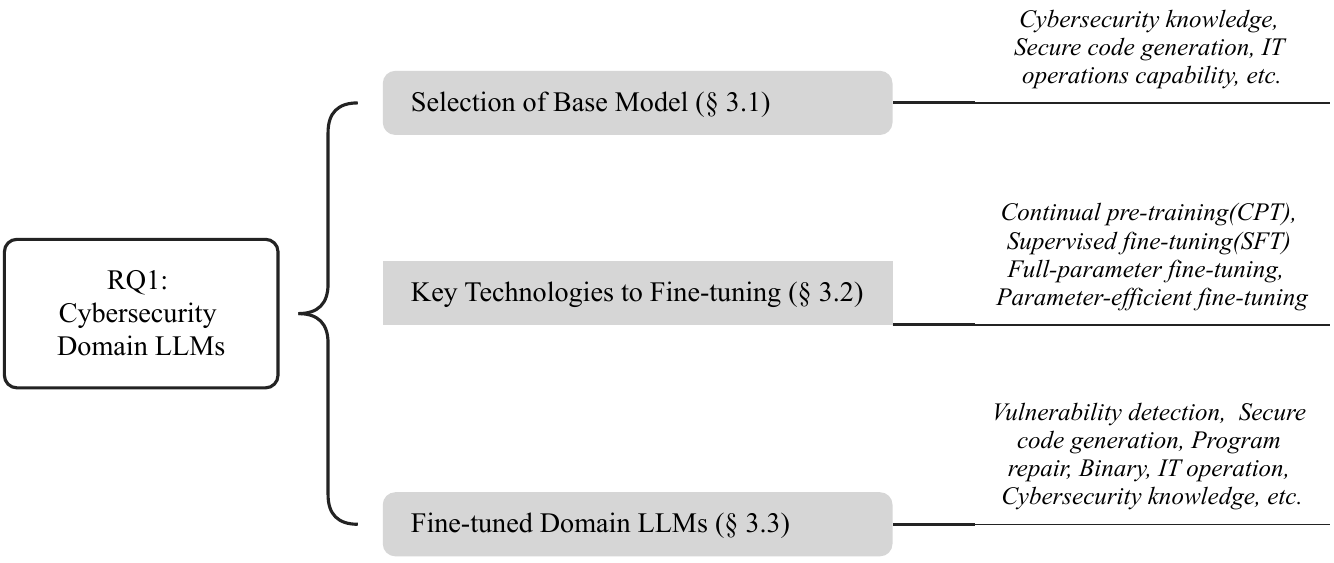}
    \caption{\textbf{An overview of RQ1.}}
    \label{fig:q1}
\end{figure*}

\subsection{Selection of Base Model for Constructing Domain LLM by Evaluating Cybersecurity Capabilities}
\label{subsec:evaluation}

It is challenging to train a cybersecurity LLM from scratch. The general practice is to choose a general-purpose LLM as the base model and then fine-tune it. However, how do we select the appropriate base model among various LLMs? \textbf{The basic idea is to choose the LLM with strong cybersecurity capabilities or those that perform well in specific security tasks}. Such models are better at understanding and addressing security-related problems. Existing evaluation of LLM cybersecurity capabilities can be divided into three categories: cybersecurity knowledge, secure code generation, and IT operations capability.

\textbf{Cybersecurity knowledge} evaluation focuses on evaluating the model's understanding of cybersecurity concepts and its ability to provide accurate information on security threats and mitigation strategies. 
CyberBench~\cite{liucyberbench} is a domain-specific, multi-task benchmarking tool for evaluating LLMs' capabilities in cybersecurity tasks. It offers a generic and consistent approach that alleviates the limitations previously encountered in evaluating LLMs in this domain. 
SecEval~\cite{li2023seceval} is designed to evaluate cybersecurity knowledge in LLMs. It provides more than 2,000 multiple-choice questions in 9 domains: \textit{Software Security, Application Security, System Security, Web Security, Cryptography, Memory Safety, Network Security, and PenTest}. By facilitating the evaluation of ten state-of-the-art foundational models, this study provides new insights into their performance in the cybersecurity domain. 
By combining expert knowledge with the collaboration of LLMs, \cite{tihanyi2024cybermetric} create the CyberMetric benchmark dataset, which contains 10,000 questions and is designed to evaluate the cybersecurity knowledge of various LLMs within the cybersecurity field. 
SecQA~\cite{liu2023secqa} is a dataset of multiple-choice questions generated by GPT-4 based on the textbook ``Computer Systems Security: Planning for Success,'' which is designed specifically to assess LLMs’ understanding and application of security principles. SecQA provides questions at two tiers of complexity, which can not only serve as an assessment tool but also facilitate the advancement of LLM applications in environments that require a high level of security awareness.
In addition, SECURE \cite{bhusal2024secure} is a benchmark designed to assess LLMs' performance in realistic cybersecurity scenarios, which includes 6 datasets to evaluate the capabilities of knowledge extraction, understanding, and reasoning in the Industrial Control System scenarios.

\textbf{Secure code generation} tests the model's capability to generate code that is not only functional but also adheres to security best practices, aiming to minimize vulnerabilities. 
CyberSecEval~\cite{bhatt2023purple} is a security coding benchmark that aims at assessing the potential security risks and tendencies to facilitate cyber attacks when LLMs generate code. By evaluating seven models including Llama 2, Code Llama, and OpenAI’s GPT, CyberSecEval effectively pinpoints key cybersecurity risks and provides practical insights for model improvement. 
LLMSecEval~\cite{tony2023llmseceval} is a dataset of 150 natural language prompts based on the narrative descriptions of various vulnerabilities that appear in MITRE's Top 25 Common Weakness Enumeration (CWE) rankings. LLMSecEval evaluates the security of LLM-generated code by comparing it to secure implementation examples for each prompt. 
SecurityEval~\cite{10.1145/3549035.3561184} focuses on the security evaluation of code generation models to prevent the creation of vulnerable code and thus avoid potential misuse by developers. This dataset includes 130 samples covering 75 types of vulnerabilities mapped to CWE. 
PythonSecurityEval~\cite{alrashedy2024llms} is a real-world dataset collected from actual scenarios on Stack Overflow, which is designed to evaluate LLMs' ability to generate secure Python code and their capacity to fix security vulnerabilities.
DebugBench~\cite{tian2024debugbench} has 4,253 instances covering four major bug categories and 18 minor types in C++, Java, and Python. This comprehensive evaluation clarifies the advantages and disadvantages of LLMs in automated debugging, which marks a major step in understanding their applicability and restraint in practical coding scenarios. 
EvilInstructCoder \cite{hossen2024assessing} is designed to assess the cybersecurity vulnerabilities of instruction-tuned Code LLMs to adversarial attacks. By incorporating practical threat models to reflect real-world adversaries with varying capabilities and evaluating the exploitability of instruction-tuned Code LLMs under these diverse adversarial attack scenarios.
Eyeballvul~\cite{chauvin2024eyeballvul} is a benchmark designed to test the vulnerability detection capabilities of language models at scale, which have contained 24,000$+$ vulnerabilities across 6,000$+$ revisions and 5,000$+$ repositories.

\textbf{IT operations capability} is used to evaluate the model's proficiency in managing and securing IT infrastructures, including awareness of security situations, security threat analysis, and incident response. 
NetEval~\cite{miao2023empirical} is an evaluation set designed to measure the common knowledge and reasoning abilities of LLMs in NetOps. This set contains 5,732 questions related to NetOps, covering five different NetOps subdomains. With NetEval, researchers systematically evaluate the NetOps capabilities of 26 publicly available LLMs. 
Additionally, OpsEval~\cite{liu2024opseval} contains 7184 multi-choice questions and 1736 question-answering formats in English and Chinese. It aims to analyze the root cause of faults, operational script generation, and alert information summarization to evaluate the performance of LLMs in IT operational tasks comprehensively. 
\cite{donadel2024can} develop a thorough framework for evaluating LLMs' capabilities in various network-related tasks and conduct an exhaustive study on LLMs' comprehension of computer networks.

In addition, NYU CTF Dataset~\cite{shao2024nyu} and Cybench~\cite{zhang2024cybench} are used to assess LLMs capacity to solve Capture the Flag (CTF) challenges in cybersecurity, aiming to improve the efficiency of LLMs in interactive cybersecurity tasks and automated task planning.
AttackER~\cite{deka2024attacker} consists of 18 distinct types of entities, which can be used for entity recognition in attack attribution and investigation tasks, revealing the potential of LLMs capabilities to improve the named entity recognition tasks in cybersecurity datasets.
SEvenLLM~\cite{ji2024sevenllm} is a framework to benchmark, elicit, and improve cybersecurity incident analysis and response abilities in LLMs for security events.

The evaluation of LLMs' cybersecurity capabilities not only guides the basic model during fine-tuning but also demonstrates that general LLMs have certain cybersecurity capabilities. This supports the feasibility of directly using LLMs (without fine-tuning) to aid cybersecurity applications, as discussed in section~\ref{sec:q2}. Furthermore, these studies help researchers and developers recognize the limitations of LLMs in the field of cybersecurity, thereby providing the direction for artificial intelligence toward higher standards and more professional security development.

\subsection{Key Technologies in Constructing Domain LLMs}

LLMs have demonstrated remarkable language understanding and generation capabilities by leveraging the transformer architecture and self-supervised pre-training strategies~\cite{vaswani2017attention,radford2018improving,brown2020language}. However, developing a specialized LLM for cybersecurity from scratch requires a lot of computational resources, which is impractical for most research teams. Fortunately, existing general LLMs have acquired extensive knowledge and demonstrated excellent generalization capabilities \cite{touvron2023llama,touvron2023llama2,yang2023baichuan,jiang2024mixtral}. \textbf{By combining these pre-trained LLMs with domain datasets for training, we can adopt a more efficient approach to enhance the model's cybersecurity capabilities.} This approach not only significantly reduces the computational demands of pre-training, but also maximizes the use of the knowledge that LLMs have learned. Thereby, the model can understand and perform cybersecurity-related tasks, such as automated threat detection, vulnerability identification, and security policy recommendations.

\begin{figure}[htbp]
    \centering 
    \includegraphics[width=0.6\textwidth]{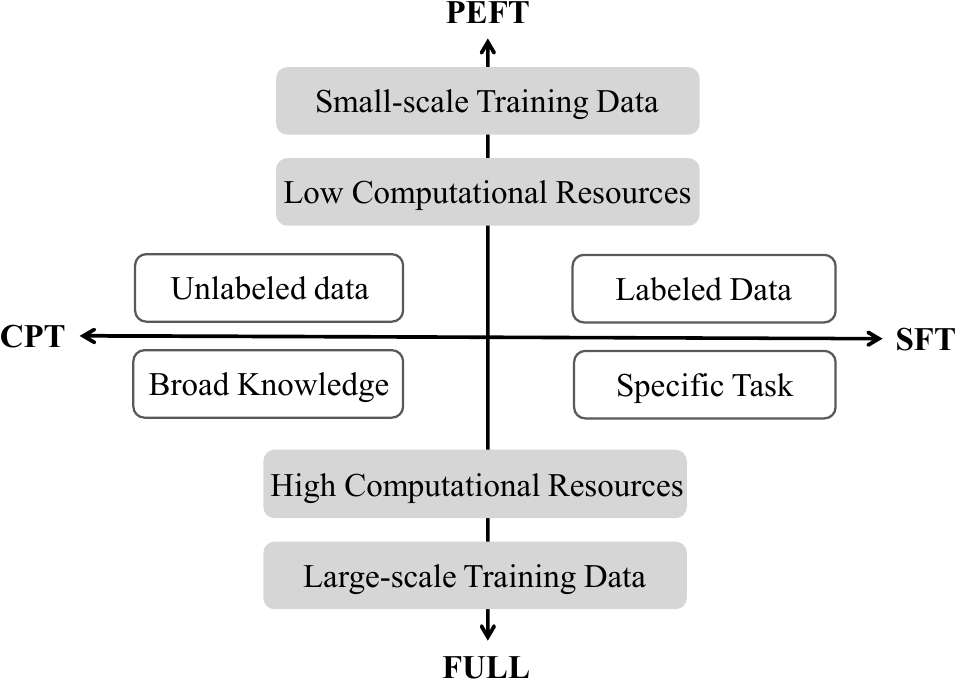}
    \caption{\textbf{Comparison of Domain LLM Training Approaches}. 
    \textit{CPT} and \textit{SFT} offer methods to enhance domain-specific performance based on existing LLMs, while \textit{FULL} parameter training and \textit{PEFT} represent different technical pathways within these training processes.}
    \label{fig:training}
\end{figure}

To apply general LLMs to cybersecurity, researchers mainly employ two approaches: continual pre-training (CPT) and supervised fine-tuning (SFT).

\textbf{Continual pre-training} involves further training of pre-trained LLMs using a large amount of unlabeled domain-specific data~\cite{yıldız2024investigating,zhang2024leveraging,wu2024continual,ibrahim2024simple}. This method aims to improve the model's understanding and application of domain knowledge, significantly improving its broad applicability within the cybersecurity field. CPT is based on the core assumption that even after extensive pre-training, the model still has the potential for further enhancement, especially in specific domains or tasks. The process usually involves several key steps: first, select a dataset that can appropriately represent the characteristics of the target domain; second, determine the strategy for continuous pre-training; and finally, perform pre-training and adjust the model architecture or optimization algorithm as needed to adapt to the new training objectives.

\textbf{Supervised Fine-Tuning} uses labeled domain-specific data for training, enabling direct optimization of the model's performance on specific cybersecurity tasks~\cite{zhang2023instruction,dong2023abilities}. Compared to CPT, SFT focuses on improving the performance of a specific task. In SFT, the model weights are refined via gradients calculated from a task-specific loss function. This function quantifies the deviation between the model's predictions and the actual labels, thus promoting the learning of task-oriented patterns. SFT relies on the utilization of high-quality, human-annotated data, which is a collection of prompts and their corresponding responses. SFT is important for LLMs such as ChatGPT, which are designed to follow user instructions and focus on specific tasks in context. This specific type of fine-tuning is also referred to as instruction fine-tuning.

In the context of CPT and SFT, researchers have the option of employing either full-parameter fine-tuning or parameter-efficient fine-tuning (PEFT). 

\textbf{Full-parameter fine-tuning} is a classical approach that adjusts all parameters of the model during training. This allows the model to fully adapt and specialize to the target domain. By optimizing all parameters, the model can achieve optimal performance for specific tasks or datasets. However, full parameter updates require considerable computing power and time, posing challenges in efficiency and scalability, especially as the number of LLM parameters continues to increase.

Conversely, \textbf{PEFT} methods fine-tune only a small number of model parameters or additional parameters while freezing most parameters of the pre-trained LLMs, which greatly reduces the computational costs. It also helps in portability, and users can fine-tune the model using PEFT methods to obtain tiny checkpoints of only a few MB in size. In summary, PEFT methods are favored because they enable users to obtain comparable performance to full fine-tuning while having only a small number of trainable parameters. There are several PEFT methods, such as adapter tuning, prefix tuning, prompt tuning, LoRA, QLoRA, and so on:
 
\textit{Adapter tuning}~\cite{he2021effectiveness} inserts adapters after the multi-head attention and feed-forward layers in the transformer architecture, which updates only the parameters in the adapter during fine-tuning while keeping the rest of the model parameters frozen. \textit{P-tuning}~\cite{liu2023gpt} automatically learns optimal task-specific prompt embeddings by introducing trainable prompt tokens, eliminating the need for manual prompt design and potentially improving performance with the addition of anchor tokens. \textit{Prefix tuning}~\cite{liu2021p} keeps the language model parameters frozen and optimizes small, continuous, task-specific vectors called prefixes. \textit{Prompt tuning}~\cite{lester2021power} fine-tunes for specific tasks through learning soft prompts by backpropagating and merging labeled examples.
\textit{LoRA}~\cite{hu2021lora} is a small trainable submodule that can be inserted into the transformer architecture. It freezes the pre-trained model weights and inserts a trainable low-rank decomposition matrix into each layer of the model, reducing the number of trainable parameters for downstream tasks. After training, the matrix parameters are combined with the original LLM. \textit{QLoRA}~\cite{dettmers2024qlora} is a further optimization of LoRA, which carries out gradient backpropagation to a low-rank adapter with a frozen 4-bit quantized pre-trained language model, reducing the memory requirement for fine-tuning while being almost comparable to full fine-tuning.

By integrating these techniques, researchers can select appropriate methods to construct LLMs tailored to the specific needs of the cybersecurity domain, as shown in Figure~\ref{fig:training}. Furthermore, emerging technologies also provide insights for the construction of cybersecurity LLMs. For example, model editing techniques~\cite{yao2023editing,zhang2024comprehensive} can precisely modify LLMs to incorporate cybersecurity knowledge without negatively affecting other knowledge. Prompt engineering~\cite{bozkurt2023generative,ye2023prompt,sahoo2024systematic}, by designing effective prompts to guide LLMs towards desired outputs, can alleviate the bottleneck of training data and resources required for constructing cybersecurity LLMs.

\subsection{Fine-tuned Domain LLMs for Cybersecurity}
\label{subsec:model}

\textbf{The researchers have used the above techniques and base models to customize LLMs to address specific problems in the field of cybersecurity.} These efforts highlight the great potential of integrating domain-specific knowledge to enhance the capabilities of LLMs, especially for key applications including vulnerability detection, fault Localization, program repair, and so on.

\textbf{Vulnerability detection} involves identifying and classifying potential security vulnerabilities in software code.
\cite{shestov2024finetuning} fine-tunes WizardCoder~\cite{luo2023wizardcoder} with LoRA specifically for vulnerability detection, focusing on the binary classification of whether Java functions contain vulnerabilities. 
\cite{ferrag2023securefalcon} performs partial parameters fine-tuning on FalconLLM~\cite{almazrouei2023falcon} using C code samples to obtain SecureFalcon, which can distinguish between vulnerable and non-vulnerable samples with a detection accuracy of up to 96\%, and further proposes a method for repairing vulnerabilities using FalconLLM. 
\cite{yang2024large} introduces a new fault localization method based on the language model, named LLMAO. LLMAO adds bidirectional adapter layers on CodeGen~\cite{nijkamp2023codegen,nijkamp2023codegen2}, enabling the model to learn bidirectional representations of codes and predict the probability of defects in code lines.
Detect Llama \cite{Ince_2024} is fine-tuned on Code-Llama with 17k dataset, outperforming GPT-4 in smart contract vulnerability detection.

\textbf{Secure code generate} via LLMs aims to improve the security of automatically generated code by mitigating vulnerability risks. 
\cite{storhaug2023efficient} proposes a new approach called vulnerability-constrained decoding, which integrates vulnerability tags during model training. By avoiding generating code with these labels, the model significantly reduces the generation of vulnerable code. Fine-tuning on GPT-J~\cite{mesh-transformer-jax} shows a notable reduction in vulnerabilities in the generated code.
\cite{he2024instruction} focuses on improving the security of code generation by LLMs via instruction tuning. They convert CodeLlama~\cite{roziere2023code} to SafeCoder using supervised fine-tuning on a dataset containing both secure and insecure programs. This approach achieves significant security improvements (approximately 30\%) across various popular LLMs and datasets while remaining practical.

\textbf{Automated program repair} aims to automatically fix software bugs without human intervention. 
\cite{silva2024repairllama} proposes a new program repair approach called RepairLLaMA, which significantly improves LLMs' program repair capabilities by applying LoRA fine-tuning to CodeLlama. It outperforms GPT-4 on the Java benchmarks Defects4J and HumanEval-Java. 
\cite{li2024exploring} first creates an instruction dataset APR-INSTRUCTION by using prompt engineering, then fine-tunes LLMs using four different PEFT methods based on this data to improve the model's automated program repair capabilities.

\textbf{Binary} is the most basic form of computer code, it is important to learn what it means and how to use it.
\cite{jiang2023nova} demonstrates the benefits of LLMs for binary analysis. They continually train StarCoder~\cite{li2023starcoder,lozhkov2024starcoder} on specialized binary code corpus and new tasks, leading to the development of Nova and Nova$^+$. After SFT, the enhanced LLMs effectively address specific tasks such as binary code similarity detection, binary code translation, and binary code recovery. 

\textbf{IT operations} manage routine tasks and activities to keep the infrastructure running for other services.
\cite{guo2023owl} describes a specialized LLM for IT operations, named Owl, which is supervised fine-tuned of Llama on the collected Owl-Instruct dataset. Owl outperforms existing models in IT-related tasks and demonstrates effective generalization capabilities on the Owl-Bench benchmark. 

\textbf{Cybersecurity knowledge assistants} help to improve users' security awareness and assist users in defending against cyber attacks through interaction with users.
\cite{hackmentor2023} proposes Hackmentor, a cybersecurity knowledge assistant. They develop a dataset of cybersecurity instructions and conversations and train Hackmentor using LoRA by fine-tuning on Llama and Vicuna~\cite{vicuna2023}. 
CyberPal~\cite{levi2024cyberpal} is fine-tuned using SecKnowledge, a domain knowledge-driven cybersecurity instruction dataset, to build a security-specialized LLM capable of answering and following complex security-related instructions.
This demonstrates the potential of LLMs in cybersecurity applications.

In addition to enhancing the cybersecurity capabilities of general LLMs through SFT and CPT, \textbf{specialized security-oriented LLMs can be developed by leveraging innovative model architectures and proprietary large-scale datasets for independent pretraining}. The Machine Language Model (MLM) is a large model designed for the machine language domain, utilizing an innovative architecture to align multimodal data across machine language, natural language, and source code \cite{liu2018alphadiff, wang2022jtrans, wang2024clap}. This approach not only addresses the limitations of existing LLMs in comprehending machine language but also introduces transformative advancements in software reverse engineering and software security detection. 
TrafficFormer~\cite{zhoutrafficformer} is an efficient pre-training model designed for traffic data. Given the characteristics of traffic data, it introduces fine-grained multi-classification tasks in the pre-training stage to enhance the representation of traffic data; in the fine-tuning stage, it uses the random initialization characteristics of the field to propose a traffic data enhancement method to help the traffic model focus on key information. In this way, the accuracy of the model's traffic detection and protocol understanding is improved.
These developments pave the way for novel research directions in the field of cybersecurity.

\begin{tcolorbox}[
  boxrule=0.5pt,
]

Answer to Q1: For researchers, it is feasible to construct the domain LLM by fine-tuning a general LLM with cybersecurity data using methods such as CPT and SFT, and the implementation technique depends on the specific application scenario, resource availability, and the expected level of performance improvement.

\end{tcolorbox}

\section{RQ2: What are the potential applications of LLMs in cybersecurity?}
\label{sec:q2}

This section introduces the application of LLMs in various cybersecurity tasks, encompassing offline defense (\textit{e.g.}, threat intelligence), online defense (\textit{e.g.}, vulnerability detection, malware detection, and anomaly detection), software testing (\textit{e.g.}, fuzz and program repair), attack assistance (\textit{e.g.}, LLM assisted attack), source code generation and analysis (\textit{e.g.}, (in)secure code generation), and other security-related applications (\textit{e.g.}, honeypot, botnet, SoC security, etc.). By reviewing the key advancements in each topic, this paper aims to offer \textbf{a comprehensive perspective on the evolution of the cybersecurity landscape driven by LLMs integration}.

\begin{figure*}[htbp]
    \centering 
    \includegraphics[width=\textwidth]{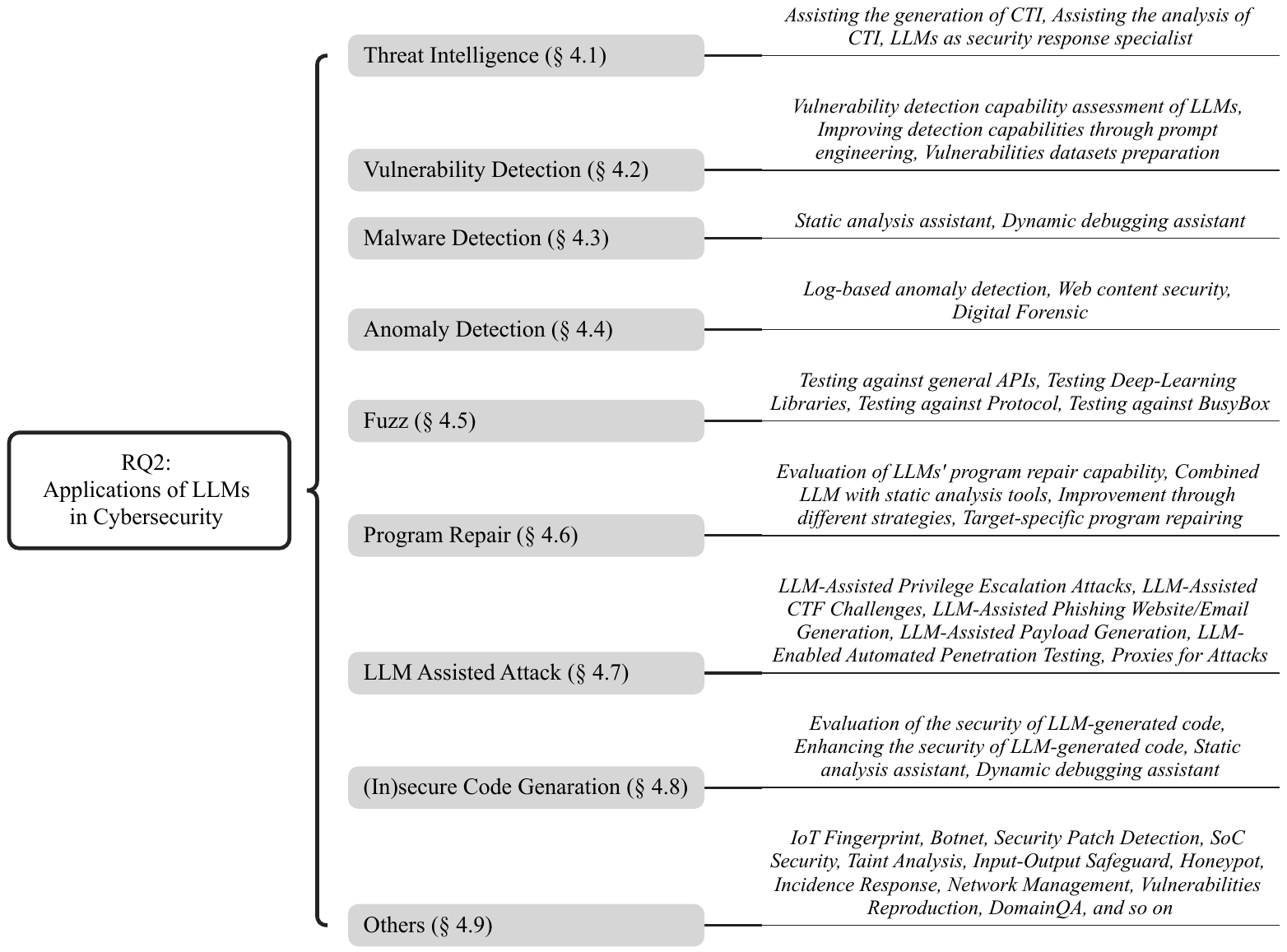}
    \caption{\textbf{An overview of RQ2.}}
    \label{fig:q2}
\end{figure*}

\subsection{Threat Intelligence}
\label{subsec:Threat Intelligence}

Since LLMs have shown excellent analysis and summarization capabilities in natural language processing tasks, \cite{clairoux2024use} assesses the performance of an LLM system built on the GPT to extract CTI information, highlight the relevance of using LLMs for CTI. More researchers have used LLMs to assist in the generation and analysis of cyber threat intelligence (CTI).

\cite{mitra2024localintel} introduces a framework known as LocalIntel, which aims to provide users with reliable threat intelligence by allowing LLMs to summarize knowledge after querying global and local knowledge databases. Global knowledge mainly refers to well-documented reports on cybersecurity threats from CWE and CVE, while local knowledge is customized by the organization for practical purposes to supplement global knowledge. \cite{perrina2023agir} also conducts similar work to extract security knowledge from a wide range of knowledge bases and automatically generate reports using LLMs. 
A few similar efforts are as follows.

\cite{fayyazi2023uses} employs LLM to generate descriptions of cyber attacks and fine-tune the model using information collected from ATT\&CK and CAPEC. Then, they compare the performance of the fine-tuned LLMs with the directly used LLMs (GPT-3.5) in describing attacks.
In another work, \cite{fayyazi2024advancing} studies the application of LLMs in cybersecurity to explain and summarize cyberattack Tactics, Techniques, and Procedures (TTPs) from the MITRE ATT\&CK framework. It compares the effectiveness of encoder-only and decoder-only models for TTP analysis and introduces Retrieval Augmented Generation (RAG) to enhance decoder-only models without fine-tuning. The study finds that RAG significantly improves the explanation of TTPs by providing relevant context, highlighting the potential of LLMs in threat intelligence. %
LMCloudHunter~\cite{schwartz2024llmcloudhunter} leverages LLMs to automatically generate generic signature detection rule candidates from textual and visual OSCTI data.
\cite{10.1145/3605770.3625214} discusses the capability of LLMs to automatically analyze and summarize software supply chain security vulnerabilities. They evaluate LLMs' performance in replicating manual assessments of 69 faults, focusing on classification accuracy. The results show that LLMs show good potential, especially when the data is comprehensive, but still cannot replace human analysts in this specific field. %
\cite{shafee2024evaluation} evaluates the performance of various LLMs in the field of threat intelligence, including ChatGPT, GPT4all, Dolly, etc. The study examines the capabilities of these chatbots in binary classification and named entity recognition (NER) tasks using a Twitter-based open-source intelligence (OSINT) dataset. While the LLMs demonstrate promising results in binary classification, their effectiveness in NER for cybersecurity entity recognition is limited, which highlights the need for further development of LLM technology to enhance CTI applications.

Specifically for digital forensics, \cite{michelet2023chatgpt} proposes a method to automate the generation of reports. They study the structure of forensic reports to identify common sections and assess the feasibility of LLMs in generating these sections. Through a case study approach, the article evaluates the strengths and limitations of LLMs in creating different sections of forensic reports.

Given that most threat intelligence providers offer information in an unstructured format, 
\cite{siracusano2023time} and \cite{hu4671345llm} propose innovative solutions to the common problem of extracting useful information from unstructured data. The former designs a framework named aCTIon, which includes downloading and parsing raw reports, extracting useful information with LLM, and exporting structured reports following STIX \cite{barnum2012standardizing} standard. The latter constructs the knowledge graph of unstructured threat intelligence and fine-tunes LLMs to automate information extraction tasks.
Also, by leveraging the capabilities of LLMs in instruction prompting and in-context learning, \cite{zhang2024attackg} propose a fully automatic LLM-based framework, AttacKG, which comprises four consecutive modules: rewriter, parser, identifier, and summarizer, to construct attack knowledge graphs from CTI.
\cite{fieblinger2024actionable} explore the application of open-source LLMs for extracting meaningful triples from CTI texts. Then, the extracted data is utilized to construct a knowledge graph, offering a structured and queryable representation of threat intelligence.

In addition to extracting valuable information from large amounts of text, report deduplication is also an important research focus in this field. 
\cite{zhang2023cupid} uses LLMs to alleviate the problem of bug report deduplication. They leverage LLMs as an intermediate step to improve the performance of REP~\cite{6100061} (a traditional method of measuring the similarity between bug reports) by identifying keywords, thereby improving its effectiveness.

There are also studies that attempt to use LLMs as experienced security response experts. 
\cite{lin2023hwv2wmap} uses LLMs as suggestion providers to mitigate vulnerabilities through prompt engineering. They design a system that is able to retrieve relative CVE \& CWE information after the user enters a vulnerability description. LLMs' mitigation suggestions are a subcomponent of the system.
\cite{kaheh2023cyber} believes that LLMs are not only question-answering assistants with expertise but also able to perform actions based on the user's description (\emph{e.g.}, instructing the host's intrusion detection system to block a specific IP).
To enhance strategic reasoning in cybersecurity, \cite{jin2024crimson} introduces Crimson, a system that uses LLMs to associate CVEs with MITRE ATT\&CK techniques to improve threat prediction and defense. The core concept is the Retrieval-Aware Training (RAT) process, which refines LLMs to generate accurate cybersecurity strategies, thereby significantly reducing errors and hallucinations. By integrating real-time data retrieval and domain-specific fine-tuning, Crimson enhances the models' interpretability and strategic consistency, providing a proactive approach to cybersecurity threat intelligence.
\cite{tseng2024using} develop an AI agent designed to automate the labor-intensive and repetitive tasks associated with analyzing CTI reports. By leveraging the advanced capabilities of LLMs, the AI agent can accurately extract important information from large volumes of text and generate Regex to help SOC analysts accelerate the process of establishing correlation rules.
\cite{rajapaksha2024rag} introduces a QA model based on Retrieval Augmented Generation (RAG) techniques together with LLMs and provides answers to the users' queries based on the knowledge base that contains curated information about cyber-attacks investigations and attribution or on outside resources provided by the users.

Considering the quality assessment of threat intelligence provided by intelligence platforms, \cite{wu2024kgv} propose a novel CTI quality assessment framework that combines knowledge graphs and LLMs. In this verifier, LLMs automatically extract OSCTI key claims to be verified and utilize a knowledge graph consisting of paragraphs for fact-checking. This significantly improves the performance of LLMs in intelligence quality assessment.

\subsection{Vulnerability Detection}
\label{subsec:Vulnerabity Detection}

This section provides an overview of the main studies on vulnerability detection using LLMs. Through these studies, we aim to shed light on the progress, challenges, and future directions of leveraging LLMs to enhance cyber security.

\textit{(In this section, we blur the concepts of "vulnerability" and "software defect")}

\textbf{Whether LLMs have the ability to detect vulnerabilities?} 
The following papers conduct preliminary studies on this question.
Although their results may vary due to some unknown reasons (\emph{e.g.}, they may use different datasets), in general, they all show that LLMs are promising for vulnerability detection~\cite{zhou2024large,tamberg2024harnessinglargelanguagemodels,zhou2024comparisonstaticapplicationsecurity,mahyari2024harnessingpowerllmssource,mao2024effectivelydetectingexplainingvulnerabilities}.

\cite{cheshkov2023evaluation} initially evaluates whether GPT-3 and GPT-3.5 could identify some known CWE vulnerabilities in Java code. The results show that the application effect in vulnerability detection tasks is not good and needs further improvement and research.
In another work, \cite{10301302} uses LLMs (including GPT-3.5, CodeGen, and GPT-4) to analyze several common vulnerabilities (\emph{e.g.}, SQL injection, overflow). The conclusion confirms that LLMs do have the ability to detect vulnerabilities, but the false positive rate is high.
However, \cite{omar2023detecting} fine-tunes GPT on various vulnerable code benchmarks to detect software vulnerabilities and achieve good performances. %
Similarly, \cite{khare2023understanding} concludes that LLMs are generally able to perform better vulnerability detection than existing static analysis and deep learning-based tools. With carefully designed prompts, desirable results can be obtained on synthetic datasets, but performance degrades on more challenging real-world datasets. %
\cite{jensen2024software} compares the performance of a wide range of open-source and proprietary models with Python code snippets in assisting vulnerability discovery. Their research suggests that LLMs can be effectively used to enhance the efficiency and quality of code reviews, particularly in detecting security issues within software code. %
\cite{shestov2024finetuning} fine-tunes WizardCoder for vulnerability detection and investigate whether the encountered performance limit is due to the limited capacity of CodeBERT-like models. Their results suggest that this is indeed the case and that LLMs have great potential for application in vulnerability detection.
\cite{li2023hitchhikers} presents LLift, a framework that leverages LLMs to assist static program analysis, specifically for detecting use-before-initialization (UBI) defects. LLift interacts with static analysis tools and LLMs, demonstrating 50\% accuracy in real-world scenarios and identifying 13 previously unknown UBI bugs in the Linux kernel. %
\cite{kouliaridis2024assessingeffectivenessllmsandroid} assess the ability of various LLMs to detect Android code vulnerabilities listed in the latest Open Worldwide Application Security Project (OWASP) Mobile Top 10. While the reported findings regarding code vulnerability analysis show promise, they also reveal significant discrepancies among the different LLMs.
Moreover, \cite{guo2024outside} thoroughly analyzes the capabilities of LLMs in detecting vulnerabilities within source code by testing the models beyond their usual applications. It also paves the way for LLM-based vulnerability detection by addressing two key aspects: model training and dataset curation

\textbf{Improving detection capabilities through different strategies.} 
Instead of directly providing code to LLM and asking it to answer, many researchers would adopt various strategies in advance. They believe that simply providing code is not enough and that the code needs to be further preprocessed or more information needs to be provided to LLMs for vulnerability reasoning.

\cite{wang2023defecthunter} proposes a code sequence embedding (CSE) that combines the AST, DFG, and CFG of the code as input to the model. Then, the model captures the semantic information with the help of conformer mechanism \cite{gulati2020conformer}, an improved architecture of Transformer.
\cite{zhang2023promptenhanced} not only provides the code to GPT but also provides the API call sequence and data flow diagrams. 
\cite{bakhshandeh2023using} conducts a similar experiment to compare the performance of the model when different levels of information are given, including asking for the vulnerability point directly, giving some CWE information, and telling LLMs what vulnerabilities are in the code. 
\cite{mathews2024llbezpeky} focuses on Android platform vulnerabilities and compares the performance of LLMs on three conditions: asking LLMs to find vulnerabilities directly, providing vulnerability summaries before asking and granting LLMs permission to request any file in the APK after providing the APK core (AndroidManifest.xml and MainActivity.java).
\cite{lee2024staticdetectionfilesystemvulnerabilities} focuses on the security of Android systems against filesystem vulnerabilities. They present PathSentinel, which leverages LLMs to generate targeted exploit code based on the identified vulnerabilities and generated input payloads, reducing the engineering effort required for writing test applications.
DLAP~\cite{yang2025dlap} combines the advantages of deep learning models for specific tasks and LLM's powerful general understanding ability, and achieves excellent vulnerability detection performance.
\cite{wang2024anvilanomalybasedvulnerabilityidentification} reframes vulnerability detection as an anomaly detection task by viewing vulnerable code as an anomaly within the LLM's predicted code distribution. This approach frees the model from the need for labeled data, allowing it to learn a representation of vulnerable code. Ultimately, it results in a detector that identifies software vulnerabilities at the line-level granularity.

There are some studies that use retrieval-augmented generation (RAG) based on additional knowledge bases to facilitate LLM for vulnerability detection.
\cite{Yang_2024} explores three different strategies for augmenting both single and multi-statement vulnerabilities using LLMs: Mutation, Injection, and Extension. These strategies potentially alleviate the shortage of data.
\cite{du2024vulragenhancingllmbasedvulnerability} proposed Vul-RAG, which leverages knowledge-level RAG framework to detect vulnerability. And the vulnerability knowledge generated by Vul-RAG can serve as high-quality explanations to further improve the manual detection accuracy.

In addition to the above efforts, researchers have also proposed many innovative ideas to improve the vulnerability detection ability of LLMs. 
\cite{hu2023large} proposes an innovative two-stage framework named GPTLENS, which includes two adversarial agent roles: auditor and critic. The auditor performs during the generation phase and its main goal is to identify potential vulnerabilities in the smart contract. In contrast, the critic works during the identification phase its main goal is to evaluate the vulnerabilities generated by the auditor.
\cite{10381286} uses traditional algorithms (TF-IDF and BM25) to match the code under analysis with the code in the vulnerability corpus to determine similarity. The code under analysis is presented to LLMs together with similar corpus entries. Based on in-context learning, LLMs can more accurately determine whether the code contains the identified vulnerability type.
\cite{sun2023gptscan} specifically focuses on vulnerability detection in smart contracts and introduce a tool called GPTScan. GPTScan first parses the smart contract project to determine the reachability of the functions, retaining only those that may have vulnerabilities. Subsequently, GPTScan uses GPT to match candidate functions with predefined vulnerability types. Finally, GPTScan asks GPT to confirm the vulnerability. %
VulLLM~\cite{du2024generalization} combines multi-task learning with LLMs, introducing two auxiliary tasks—vulnerability localization and vulnerability explanation—in addition to the primary vulnerability detection task. This approach enhances the model's ability to understand the root causes of code vulnerabilities, thereby improving its generalization capabilities.

To improve LLM's ability to reason about vulnerabilities, \cite{sun2024llm4vuln} proposes LLM4Vuln, which separates the vulnerability reasoning capabilities of LLMs from others (\emph{e.g.}, proactively seeking more information, employing relevant vulnerability knowledge, and following instructions to output structured results). They allow LLMs to request additional contextual information about the target code.
Moreover, they conclude that the more information input to LLMs is not the better. Too much information such as full vulnerabilities report, and a large amount of invocation context, may lead to distractions.
\cite{mao2024multirole} proposes a new method called MuCoLD, which simulates a multi-role code review process for vulnerability detection in software. By playing different roles, such as developers and testers, LLMs participate in discussions to reach a consensus on the existence and classification of vulnerabilities.
IRIS~\cite{li2024llmassistedstaticanalysisdetecting} combines LLMs with static analysis to enable reasoning over the entire codebase. It automatically infers taint specifications and performs contextual analysis, thereby reducing reliance on human-generated specifications and manual inspection.

In addition to detecting vulnerabilities in specific programs, recent studies have attempted to use LLMs to infer lists of affected libraries from vulnerability reports. \cite{chen2023vullibgen} observes that many vulnerability reports in the national vulnerability database (NVD) either omitted affected libraries or provided incomplete or incorrect library names, increasing the risk of third-party library vulnerabilities. To address this problem, they propose VulLibGen, a method designed to detect vulnerabilities in third-party libraries. VulLibGen takes only vulnerability descriptions as input and uses the inherent knowledge of LLMs to generate a list of library names that may be affected by the reported vulnerabilities.

\cite{liu2023chatgpt} explores the application of ChatGPT for vulnerability management. They evaluate ChatGPT's capabilities in predicting security bugs, evaluating severity, repairing vulnerabilities, and verifying patch correctness. The results reveal that while ChatGPT can assist in identifying and mitigating software security threats, it needs enhancements to perform more nuanced tasks, such as vulnerability prioritization and patch validation.

\textbf{Construction of vulnerability detection datasets.} 
In addition to the methods of retraining or fine-tuning the models, the construction of the dataset is also important for vulnerability detection.

\cite{10.1145/3607199.3607242} introduces a new vulnerable source code dataset called DiverseVul, which contains 18,945 vulnerable functions (covering 150 CWEs) and 330,492 non-vulnerable functions, all written in C/C++. They also explore 11 different deep learning architectures and conclude that despite the remarkable success of LLMs, they still face challenges such as high false positive rates, low F1 scores, and difficulty in identifying complex CWEs for vulnerability detection.
\cite{gao2023far} introduces a comprehensive vulnerability benchmark dataset called VulBench, which includes high-quality data from CTF challenges and real-world applications with detailed annotations of vulnerability types and causes for each vulnerable function. 
\cite{10.1145/3617555.3617874} creates a dataset containing 112,000 vulnerable C code instances with detailed information about the specific vulnerability, including CWE number, location, and function name. Notably, all the code in this dataset is generated by GPT-3.5, which illustrates the application potential of vulnerable code synthesized by LLMs.
Source Code Processing Engine (SCoPE)~\cite{gonccalves2024scope} is a framework that incorporates strategized techniques to reduce the size and normalize C/C++ functions. Additionally, SCoPE refines the CVEFixes dataset, which can be used for fine-tuning pre-trained LLMs for software vulnerability detection.

\subsection{Malware Detection}
\label{subsec:malware}

In malware detection, LLMs can serve as both the static analysis assistant and the dynamic debugging assistant, improving the efficiency and effectiveness of the process, and making it an important part of defending against cyber threats.

\textbf{LLMs as the static analysis assistant.}
\cite{pearce2022pop} explores the application of LLMs, such as OpenAI's Codex, in the field of reverse engineering, particularly in understanding software functionality and extracting information from the code. LLMs are primarily used to analyze the functionality of C code provided by reverse engineering tools such as Ghidra. These C codes are obtained from binary files through the process of decompilation. %
Decompilation is also an important task in reverse engineering. 
\cite{tan2024llm4decompile} introduces an LLM tailored for decompilation that focuses on converting compiled machine code back into human-readable source code. They fine-tune a model called DeepSeek-Coder on a large number of C code and assembly code pairs and evaluate the performance of their work by recompiling and executing the decompiled code.
\cite{fang2023large} explores the potential and limitations of LLMs for code analysis tasks, especially when dealing with obfuscated code. In the experiments, they conduct tests that allow LLMs to generate de-obfuscated versions of code, \emph{i.e.}, to recover more readable original code from obfuscated code.

\cite{zhao2023understanding} focuses on how to improve LLM's semantic understanding of programs through fuzz testing. Their core idea is that programs with their basic units (\emph{e.g.}, functions, and subroutines) are designed to exhibit diverse behaviors and provide possible outputs given different inputs. Thus, through fuzz testing, various inputs trigger different functions of the code that can help LLMs understand the behavior and semantics of the program more thoroughly..  %
\cite{palacio2023evaluating} introduces ASTxplainer, an explainability method for LLMs in coding scenarios. It aligns token predictions with Abstract Syntax Tree (AST) nodes, enabling detailed evaluation and visualization of model predictions. ASTxplainer consists of AsC-Eval for structural performance estimation, AsC-Causal for causal analysis, and AsC-Viz for visualization. These components provide a more comprehensive explanation of how LLMs work when generating or analyzing code. %

\cite{yan2023prompt} focuses on how LLMs can be utilized to aid in dynamic analysis of malware. The core idea of the research is to use GPT-4 to generate explanatory text for each API call, and then use BERT to generate a series of API sequences to be executed based on the previous analysis. This approach can theoretically generate representations for all API calls without the need to train the dataset during the generation process. %
\cite{PPR:PPR763542} uses LLM (specifically ChatGPT) to analyze the linguistic and strategic elements of ransomware communications. By examining a range of ransomware samples, the study identifies patterns and strategies used in ransom notes, revealing the evolution of ransomware strategies characterized by sophisticated language use and psychological manipulation. 
\cite{wang2023using} also discusses the potential and challenges of LLMs in generating strategies against ransomware. %
\cite{zahan2024shifting} employs GPT-3 and GPT-4 to detect potential malware in the npm ecosystem by analyzing JavaScript packages. The study introduces SocketAI Scanner, a multi-stage workflow that utilizes iterative self-refinement, zero-shot role-playing, and chain of thought prompting techniques to enhance the model's ability to identify malicious intent within code. By comparing LLMs' performance with static analysis tools, the paper demonstrates that LLMs can effectively pinpoint malware with higher precision and lower false positive rates.

Binary malware summarization aims to automatically generate human-readable descriptions of malware behaviors from executable files, facilitating tasks like malware cracking and detection.
\cite{lu2024malsightexploringmalicioussource} introduces a novel code summarization framework, namely MALSIGHT, which can iteratively generate descriptions of binary malware by exploring malicious source code and benign pseudocode. At the same time, they construct the first malware summary dataset, MalS and MalP, to support further research.

\textbf{LLMs as the dynamic debugging assistant.}
\cite{tian2024debugbench} introduces DebugBench, a benchmark for evaluating LLMs' debugging capabilities in programming. It consists of 4253 instances across various bug categories in C++, Java, and Python. The benchmark is constructed by collecting code snippets from LeetCode, implanting bugs with GPT-4, and conducting rigorous quality assessment. %
\cite{liu2023make} addresses the challenge of automated Graphical User Interface (GUI) testing for mobile applications. They propose a novel approach called GPTDroid that formulates the GUI testing as a question and answering (Q\&A) task, where the LLM is asked to chat with the mobile apps by passing GUI page information to generate testing scripts. These scripts are executed and iterations of the application's responses are fed back to the model to guide further exploration.
\cite{ahmad2023flag} proposes an approach called FLAG to assist human debuggers in identifying and localizing security and functional bugs in code. FLAG takes a code file as input and regenerates each line in the file for comparison. It compares the original code with LLM-generated code to flag notable differences as anomalies for further inspection.  %

\subsection{Anomaly Detection }
\label{subsec:Anomaly Detection}

We investigate some methods to incorporate LLMs into cybersecurity frameworks for anomaly detection, underscoring their critical role in maintaining network integrity and safeguarding against cyber intrusions.

\textbf{Log-based anomaly detection.}
\cite{karlsen2023benchmarking} tests 60 language models fine-tuned for log analysis, including models with different architectures such as BERT, RoBERTa, DistilRoBERTa, GPT-2 and GPT-Neo. The results show that these fine-tuned models can be effectively used for log analysis, especially for domain adaptation for specific log types. %
Targeting service logs on Huawei Cloud, \cite{liu2023logbased} proposes a framework called ScaleAD, which aims to provide an accurate, lightweight, and adaptive solution for log anomaly detection in cloud systems. When ScaleAD's Trie-based Detection Agent (TDA) detects suspicious anomaly logs, it queries the LLM to validate these logs. The LLM determines whether the logs are anomalous or not by understanding the semantics of the log content and gives the corresponding confidence scores.
\cite{qi2023loggpt} proposes a log anomaly detection framework named LogGPT. This framework consists of three main components: log preprocessing, prompt construction, and response parser. The log preprocessing component filters, parses and groups raw log messages into a structured format for further analysis. The response parser extracts the output returned by ChatGPT for detailed analysis and evaluation of the detected anomalies.
\cite{han2023loggpt} performs similar work. The difference is that they fine-tune GPT-2 by introducing a Top-K reward metric, which directs the model to focus on the most relevant parts of the log sequence, thus improving the accuracy of anomaly detection.
\cite{liu2024interpretable} introduces an online log analysis method called LogPrompt. They employ LLMs to parse unstructured logs and generate reports with a specific structure. LogPrompt then utilizes chain of thought and in-context learning methods to progressively reason about log content and provide normal/abnormal judgments.
\cite{zhang2024lemur} introduces LEMUR, a cutting-edge log parsing framework that enhances log analysis with entropy sampling for efficient log clustering and semantic understanding using LLMs. LEMUR addresses the limitations of traditional parsers by discarding manual rules and focusing on semantic information. Relying on semantic understanding of LLMs, the framework accurately distinguishes between parameters and invariant tokens, leading to impressive efficiency and state-of-the-art performance in log template merging and categorization.

\textbf{Web content security.}
LLMs can assist in the detection of phishing and spam.
\cite{jamal2023improved} presents a model named Improved Phishing and Spam Detection Model (IPSDM), a fine-tuned model based on DistilBERT and RoBERTa. They emphasizes the potential of LLMs to revolutionize the field of email security and suggests that these models can be valuable tools for improving the security of information systems.
Another work also conduct spam detection with LLMs, \cite{wu2024evaluating} evaluate ChatGPT's performance in spam detection and find it outperforms BERT on a low-resource Chinese dataset but lags on a larger English dataset. The study also highlights the positive impact of increasing prompts on ChatGPT's accuracy.
\cite{nahmias2024prompted} introduces a spear-phishing detection approach utilizing LLMs to generate ``prompted contextual document vectors.'' By posing targeted questions to LLMs about email content, the method quantifies the presence of common persuasion principles, creating vectors that capture the malicious intent within spear-phishing emails. The approach utilizes the reasoning capabilities of LLMs and outperforms traditional phishing detection methods.
In addition to detecting phishing emails, there are studies on generating phishing emails using LLMs. \cite{heiding2023devising} evaluates the performance of GPT-4 in creating phishing emails and compare its effectiveness with traditional phishing methods called V-Triad method, which relys on manual design based on general rules and cognitive heuristics. They also explore the use of LLMs in detecting phishing emails, where models like GPT, Claude, PaLM, and LLaMA demonstrate strong capabilities in identifying malicious intent, sometimes surpassing human detection rates.

In addition, LLMs can be used for malicious URLs, DDoS attacks, and other cyber threat detection.
Based on the website content, \cite{vörös2023web} uses the knowledge distillation approach to detect malicious URLs. Specifically, unlabeled URLs are classified and labels are generated by a teacher model. The student model trained with this label improves accuracy with significantly fewer parameters and is therefore suitable for malicious URL detection. 
\cite{guastallaapplication} explores the potential of LLMs in detecting DDoS attacks by investigating the performance of LLMs on two datasets. For the CICIDS 2017 dataset, they fine-tune LLMs with labeled pcap files to enable traffic classification through few-shot learning. Urban IoT dataset is a real-world anonymized dataset containing 4060 IoT devices. Considering the complexity of this dataset, they fine-tune LLMs separately depending on whether the correlation of traffic between IoT devices is considered or not.
\cite{ferrag2023revolutionizing} encodes the network traffic by employing a novel encoding technique called Privacy-Preserving Fixed-Length Encoding (PPFLE). Then they train a model named SecurityBERT with these encoded data to perform a classification task on network traffic. Specifically, their model targets IoT devices to achieve efficient and accurate cyber threat detection on resource-limited IoT devices.
\cite{ziems2023explaining} studies the interpretation of decision tree models in network intrusion detection (NID) systems. They convert the path and structure data of the decision tree into text format and provide it to LLMs to generate explanations. Moreover, LLMs provide additional background knowledge to help users understand why certain features are important in categorization.
\cite{ali2023huntgpt} introduces HuntGPT, a system that integrates LLMs with traditional machine learning for anomaly detection. The system utilizes a random forest classifier trained on the KDD99 dataset to identify cyber threats. To enhance interpretability, the system employs XAI techniques such as SHAP and Lime and combines them with the GPT-3.5 conversational agent.

\textbf{Digital Forensic.}
\cite{SCANLON2023301609} assesses the applicability of ChatGPT for digital forensics. ChatGPT is used to help determine if a file has been downloaded to a PC and if the file has been executed by a specific user. In addition, ChatGPT is also used to detect browser history, Windows event logs, and interactions with cloud platform machines. %

\subsection{Fuzz}
\label{subsec:Fuzz}

Although traditional fuzzing techniques are effective in discovering software vulnerabilities, their inherent limitations can affect their efficiency and effectiveness. One significant drawback is that traditional fuzzers operate in a largely random or semi-random manner, which is time-consuming and inefficient because they may not explore all possible execution paths. Additionally, the mutated seeds are usually artificially crafted, which makes the time and labor costs high.
Although all of the above problems have been studied for many years and there are many ways to mitigate them, the emergence of LLMs provides a new way of thinking in the field of fuzz testing~\cite{jiang2024fuzzing,wang2024exploratory}.

\textbf{What are the advantages of LLMs fuzz over traditional methods?}
\cite{zhang2023does} evaluates the performance of ChatGPT in generating test cases directly (without tuning) and compare it with two traditional testing tools (\emph{i.e.}, SIEGE, and TRANSFER). Their experiments show that LLMs outperform traditional methods in generating test cases when a detailed description of the vulnerability, possible exploits, and code context are given.

There are some advantages of LLMs over traditional tools.
One of the most important factors is that LLMs lead to a shift from random mutation to guided mutation. \cite{hu2023augmenting} introduces a GPT-based seed mutator to the traditional gray-box fuzz testing, selecting seeds from a seed pool and requesting variants from ChatGPT to generate higher-quality inputs. 
Another factor is that LLMs have a strong understanding of programming languages, enabling them to perform testing tasks in multiple languages. Most traditional methods can only fuzz specific programming languages. \cite{xia2024fuzz4all} tests 6 languages code (\emph{i.e.}, C, C++, Go, SMT2, Java, and Python) with a method named Fuzz-Loop, which automatically mutates test cases based on LLMs.
Most traditional fuzz methods fail to achieve high code coverage in all codes, while LLMs have mastered the logic of code and can generate more targeted test cases for areas with low coverage. For example, 
\cite{10172800} uses Codex to generate test cases against low-coverage functions when SBST (Search-Based Software Testing, a traditional fuzz method) reaches coverage plateau. Specifically, the raw character sequences generated by the Codex are deserialized into an internal test case representation compatible with SBST to leverage its mutation operations and fitness functions.

\textbf{Specific fuzzing strategies for different testing objects.}
Depending on the test subject, the strategy should be adjusted when fuzzing with LLMs. 
For \textbf{testing against general APIs}, \cite{zhang2023understanding} investigates the effectiveness of LLMs in generating invocation code. They compare LLM-based generation with traditional program analysis methods and find that LLMs can automatically generate a large number of effective fuzzing drivers while reducing human intervention. The research introduces query strategies, iterative improvements, and the use of examples to enhance LLM performance. 
Although it's all about testing APIs, the strategy for \textbf{testing deep-learning libraries} needs to be modified. Because programs that call deep learning libraries usually have strict requirements on tensor dimensions, ignoring this would cause the fuzzer to perform meaningless tests.
\cite{10.1145/3597926.3598067} proposes TitanFuzz, a tool specifically for generating test cases for deep learning libraries. Their training corpus contains a large number of code snippets that call the DL library APIs, so that the language syntax and semantics, and complex DL API constraints can be learned to efficiently generate DL programs.
FuzzGPT \cite{deng2023large} is also about fuzzing the DL library. The difference is that FuzzGPT focuses on using historical error-triggered code snippets to guide LLMs to generate test cases.

In addition to the above research, we have collected some studies targeting other testing objects. \textbf{Testing against Protocol}. \cite{meng2024large} discusses how to find security vulnerabilities in protocol implementations in the absence of a machine-readable protocol specification. They train LLMs with massive human-readable protocol documents and ask LLMs to mutate interactive messages for protocol fuzz (\emph{e.g.}, HTTP). 
\textbf{Testing against BusyBox}. Specifically targeting BusyBox, a popular utility in Linux-based devices, \cite{asmita2024fuzzing} introduces two fuzzing methods. One is to use LLMs to generate target-specific initial seeds for fuzzing, which significantly improves the efficiency of identifying crashes and potential vulnerabilities. The other is crash reuse, which employs previously acquired crash data to streamline the testing process for new targets.

\subsection{Program Repairing}
\label{subsec:Program Repairing}

The software development lifecycle is deeply impacted by the presence of bugs, with their detection and resolution being costly. Researchers are motivated to find new ways to automatically identify and correct bugs/vulnerabilities with LLMs~\cite{zhang2024systematicliteraturereviewlarge}.

\textbf{Evaluation of existing LLMs on program repairing.}
For state-of-the-art LLMs (open-sourced or proprietary), many studies have evaluated their capabilities for program repairing.
\cite{prenner2021automatic} explores the application of OpenAI's Codex to the field of automatic program repair (APR), specifically its ability to locate and fix bugs in software. They use the QuixBugs benchmark, which includes 40 bugs in Python and Java, to evaluate the effectiveness of Codex in APR tasks. Notably, Codex outperforms numerous existing APR methods even without retraining.
\cite{sobania2023analysis} conducts similar work with the previous one. Both studies evaluate LLMs for automatic program repair on QuixBugs benchmark. In this work, ChatGPT is evaluated instead of Codex.  %
\cite{52980} discusses the application of Gemini in automating the repair of software vulnerabilities, especially for vulnerabilities found by the sanitizer tool in C/C++, Java, and Go code. The authors argue that while the success rate seems low, it has the potential to significantly reduce engineering effort over time.  %
\cite{yu2024security} evaluates the performance of three LLMs, Gemini Pro, GPT-4, and GPT-3.5, on codes with identified vulnerabilities from real-world code reviews. The findings indicate that GPT-4 outperforms the other models, but all LLMs have great potential, especially for conciseness, clarity, and accuracy of responses.
\cite{xia2022practical} selects 9 LLMs and compare them with traditional automated program repair methods, demonstrating the superior effectiveness of LLMs in this field. %

\cite{10179324} explores the potential of LLMs for zero-shot vulnerability repair in code. Through extensive experiments with various LLMs in synthetic, artifactual, and real-world security scenarios, they demonstrate that while LLMs show promise in repairing simple cases, they struggle with more complex, real-world examples. The study reveals the limitations and strengths of LLMs in cybersecurity and urges further research into the application of LLMs in program repairing.
\cite{Wu_2023} compares the capabilities of LLMs and deep learning-based APR models in fixing Java vulnerabilities. They evaluate the performance of 5 LLMs (Codex, CodeGen, CodeT5, PLBART, and InCoder), 4 fine-tuned LLMs, and 4 deep learning-based APR techniques on two real-world Java vulnerability benchmarks (\emph{i.e.}, Vul4J and VJBench). They design code transformations to address the overlapping of train and test sets faced by Codex, and create a new Java vulnerability remediation benchmark, VJBench, to better evaluate LLMs and APR techniques. %
\cite{xiang2024farpracticalfunctionlevelprogram} investigate LLM-based function-level APR, focusing on the effects of the few-shot learning mechanism and the inclusion of auxiliary repair-relevant information. The study shows that LLMs with zero-shot learning are already effective for function-level APR, but applying the few-shot learning mechanism results in varying repair performance. Additionally, they find that directly incorporating auxiliary repair-relevant information into LLMs significantly enhances function-level repair performance.

\textbf{Combined LLMs with static analysis tools.}
Instead of using LLMs alone for program repair, some studies have combined them with traditional program analysis tools to increase the efficiency of those tools.
\cite{alrashedy2024llms} proposes a new approach called Feedback-Driven Security Patching (FDSP), which passes feedback from Bandit to LLM. With the help of the static code analysis tool, LLM can generate potential solutions to address security vulnerabilities. Each suggested solution, along with the corresponding vulnerable code segment, is fed back to LLM for verification and validation.
\cite{jin2023inferfix} introduces a program repair framework called InferFix that incorporates the latest static analyzers for fixing critical security and performance vulnerabilities. Inferfix consists of two main components: a retriever and a generator. The retriever aims to search for semantically similar vulnerabilities and their associated fixes. The generator is fine-tuned on vulnerability fix data, with prompts enhanced by bug type annotations and semantically similar fixes, thereby improving the model's ability to generate effective proposals.

\textbf{Improving repair capabilities through different strategies.}
To improve the performance of LLMs on program repair tasks, researchers have proposed some methodologies.
D4C~\cite{xu2024aligningllmsflfreeprogram} is a straightforward prompting framework for APR. By aligning the output to LLMs training objective and allowing LLMs to refine the whole program without first identifying faulty statements, D4C greatly improve LLM's APR capability.
\cite{chen2023teaching} proposes an approach called SELF-DEBUGGING. Even if there is no human feedback about the correctness of the code or error messages, this method can identify the error by observing the execution results and explaining the code generated by natural language.
\cite{10298561} explores the application of Self-Consistency (an approach for improving model reasoning ability \cite{wang2023selfconsistency}) in program repair. By incorporating commit-logs as reasoning paths in few-shot prompts, Self-Consistency enables LLMs to generate diverse solutions. The most frequent solution from multiple samples is selected to improve patch accuracy. 
Similarly, VRpilot~\cite{kulsum2024casestudyllmautomated} is based on reasoning and patch validation feedback. The method uses a chain-of-thought prompt to reason about a vulnerability before generating patch candidates and iteratively refines the prompts based on feedback from external tools on previously generated patches, improving patch accuracy.
DRCodePilot~\cite{zhao2024enhancingautomatedprogramrepair} is designed to enhance GPT-4-Turbo's APR capabilities by incorporating design rationales (DR) into the prompt instruction, along with a utility feedback-based self-reflective framework. This framework prompts GPT-4 to reconsider and refine its outputs by referencing the provided patch and suggested identifiers.

Additionally, \cite{yang2024revisitingunnaturalnessautomatedprogram} introduces a novel approach that leverages the entropy of LLMs in combination with prior APR tools to enhance all stages of the APR process. By using entropy-delta for patch ranking and classification, this method can rank correct patches more effectively than state-of-the-art machine learning tools.
ThinkRepair~\cite{yin2024thinkrepairselfdirectedautomatedprogram} is an LLM-based autonomous two-stage automatic program repair framework. In the collection stage, CoT prompts guide the LLM to automatically gather various reasoning chains that form the foundation of the repair knowledge. In the repair stage, sample selection is performed for few-shot learning, with interactive feedback from the LLM. This approach significantly improves LLMs' bug fixing capability.
\cite{10.1145/3611643.3616271} proposes a program repair framework named Repilot. It starts by masking the buggy code segment and then utilizes LLMs to generate candidate patches. During the generation, Repilot consults the completion engine to prune infeasible tokens and proactively completes the code when necessary. This approach enhances the compilation rate and correctness of patches while reducing the number of invalid attempts in the generation process. %
\cite{islam2024llmpowered} introduces SecRepair, a system that uses LLMs to detect and fix code vulnerabilities in the software. It utilizes reinforcement learning with the semantic reward mechanism to improve the model's ability to generate accurate code comments and descriptions, guiding developers to address security issues.
ARJA-CLM~\cite{wang2024revisitingevolutionaryprogramrepair} integrates a multi-objective evolutionary algorithm with a code language model to fix multi-location bugs in Java projects. It does this by predicting the correct statement for masked buggy positions using the powerful code-filling capabilities of CodeLLMs.
\cite{Kong2024ContrastRepairEC} launches Contrastrepair to provide more accurate feedback by providing LLMs with contrastive test case pairs (a failing test and a passing test), thereby enhancing conversation-driven repair framework. The key insight is to minimize the difference between the generated passing test and the failing one, effectively isolating bug causes. ContrastRepair interacts with ChatGPT repeatedly to generate patches until a plausible fix is generated.
Unlike previous function-level approaches, \cite{chen2024large} investigates the performance of LLMs in repository-level program repair, which needs to consider interactions and dependencies between code that may span multiple functions or files. In this work, they propose a benchmark named RepoBugs, which includes 124 bugs from open source repositories to evaluate the performance of LLMs.

Fine-tuning is also necessary to unlock state-of-the-art performance in program repair. MORepair~\cite{yang2024multiobjectivefinetuningenhancedprogram} is a multi-objective fine-tuning approach that instructs LLMs to generate high-quality patches. It involves adapting the LLM parameters to the syntactic nuances of code transformation and specifically fine-tuning the model to understand the logical reasoning behind code changes in the training data. This fine-tuning strategy enables LLM to achieve superior performance in program repair.
\cite{defiterodominguez2024enhanced} fine-tunes LLM on datasets containing C code vulnerabilities. They specifically design a structured representation of the code and provide it to LLM, including the line number of the code that needs to be repaired, the vulnerability description (\emph{i.e.}, CWE description), and the complete source code. The output of LLM is also structured and can be directly patched, which enables the code to be repaired automatically without manual intervention. 
\cite{zhao2024repairautomatedprogramrepair} explores how LLMs can achieve excellent APR performance through process supervision and feedback. They first construct a dataset called CodeNet4Repair, which is filled with multiple repair records for supervised fine-tuning. Then,they develop a reward model that provides feedback on the fine-tuned LLM's actions, progressively optimizing its policy for better repair.
\cite{dehghan2024mergerepairexploratorystudymerging} proposes continual merging and empirically studies the capabilities of merged adapters in Code LLMs for the APR task. Specifically, task-specific adapters are first trained for the LLM, and then MergeRepair is used to merge multiple task-specific adapters, considering the order and weight of the merged adapters for better APR.

\textbf{Target-specific program repairing.}
We also investigate some studies on program repairing for some specific targets.
\cite{tol2023zeroleak} proposes a framework called ZeroLeak, which explore how LLMs can be used to automatically generate repair code to address side-channel vulnerabilities in software. ZeroLeak guides LLMs to generate patches for specific vulnerabilities through zero-shot learning. Once generated, these patches are inspected by dynamic analysis tools to ensure that they not only function correctly, but also prevent information leakage. %
\cite{paria2023divas} introduces a novel framework named DIVAS. The framework maps user-defined SoC specifications to Common Weakness Enumerations (CWEs), generates SystemVerilog Assertions (SVAs) for verification, and enforces security policies. DIVAS automates the process of vulnerability detection and policy enforcement, reducing manual effort and enhancing SoC security.  %
\cite{ahmad2023fixing} constructs a corpus of hardware security vulnerabilities and utilize LLMs to automatically remediate Verilog code containing these vulnerabilities. %
\cite{charalambous2024automatedrepairaicode} focuses on the software implementation of neural networks and related memory safety issues, including NULL pointer dereferencing, out-of-bounds access, double-free errors, and memory leaks. They propose detecting these vulnerabilities and automatically repairing them with the help of LLMs.
\cite{khang2024study} focuses on the application of LLMs(\emph{e.g.}, ChatGPT and Bard) in repairing security vulnerabilities in JavaScript programs.Using the top 25 CWEs of 2023 as a reference, they selecte JavaScript-related vulnerabilities to evaluate the accuracy of the models in generating the correct patches. Their findings highlight the potential of LLMs for JavaScript security, emphasizing the effectiveness of LLMs for programming languages used for web development. %
To convert a regular C/C++ program into its HLS-compatible counterpart (HLS-C), \cite{xu2024automatedccprogramrepair} proposes an LLM-driven program repair framework that takes standard C/C++ code as input and automatically generates the corresponding HLS-C code for synthesis, minimizing human repair effort.

\subsection{LLM Assisted Attack}
\label{subsec:LLM Assted Attack}
A report \cite{barrett2023identifying} from the workshop organized by Google on January 1, 2024 highlight the dual-use issue of Generative Artificial Intelligence (GenAI). These techniques can be used for both positive purposes and potentially for malicious attacks. In this section, we discuss current attacks with the help of LLMs in detail.

\textbf{Current status of LLM-assisted attacks.}
\cite{10127411} points out that ChatGPT has both positive and potentially negative impacts on cybersecurity. They list various types of threats to cybersecurity today, including malware attacks, phishing, and password attacks. They also mention the potential application of ChatGPT in social engineering attacks. %
\cite{10198233} also conduct similar work on the impact of generative AI in cybersecurity and privacy.
Furthermore, \cite{moskal2023llms} explores the potential of LLMs for network threat testing, particularly in supporting threat-related actions and decisions. Experimenting on virtual machines, they discuss in detail how automated attacks guided by LLMs can be launched against devices in a network. They conclude that while this work is preliminary, it demonstrates that LLMs shows strong potential for cyber threats.
For existing accessible malicious LLMs, \cite{lin2024malla} conducts a systematic study of 212 real-world Malla (malicious LLM), revealing how they spread and work in the underground market. They examine in detail the Malla ecosystem, development frameworks, exploitation techniques, and the effectiveness of Malla in generating various malicious content. They also provide insights into  how cybercriminals utilize LLMs and strategies for combating such cybercrime. %

Specifically, there are various means of executing automatic attacks with the help of LLMs.

\textbf{LLM-Enabled Automated Penetration Testing.}
\cite{deng2023pentestgpt} introduces a tool called PentestGPT designed to perform automated penetration tests. PentestGPT consists of three modules: inference, generation and parsing. Each module reflects a specific role in the penetration testing team so that the system can more realistically simulate automated penetration tests.
\cite{10.1145/3611643.3613083} also conducts a study on penetration testing with the help of LLMs. The study investigates two use cases: high-level task planning for security testing and low-level vulnerability hunting within vulnerable virtual machines. They cerate a feedback loop between LLM-generated operations and the virtual machine, allowing LLMs to analyze the state of the system to find vulnerabilities and suggest attack vectors. %
\cite{huang2024penhealtwostagellmframework} points out the importance of integrating penetration testing with vulnerability remediation into a cohesive system. They proposes PenHeal, a two-stage LLM-based framework designed to autonomously identify and mitigate security vulnerabilities. The framework integrates two LLM-enabled components: the Pentest Module, which detects multiple vulnerabilities within a system, and the Remediation Module, which recommends optimal remediation strategies.
\cite{Pratama_2024} developes CIPHER (Cybersecurity Intelligent Penetration-testing Helper for Ethical Researchers), a LLM trained using over 300 high-quality write-ups of vulnerable machines, hacking techniques, and documentation of open-source penetration testing tools. Additionally, they introduce the Findings, Action, Reasoning, and Results (FARR) Flow augmentation to enhance penetration testing write-ups, establishing a fully automated pentesting simulation benchmark tailored for LLMs.

\textbf{LLM-Assisted Automatic Full-Life-Cycle Cyberattack.}
\cite{xu2024autoattacker} proposes AUTOATTACKER, a system that leverages LLMs to automate the execution of "keystroke-operated" cyberattacks that mimic human operations. The system employs LLMs to generate precise attack commands for various techniques and environments, transforming potential manual operations into automated and efficient processes. AUTOATTACKER consists of multiple modules that interact iteratively with the LLM to construct complex attack sequences using functions such as summarization, planning, and action selection.
AURORA~\cite{wang2024sandsmansionsenablingautomatic} is another automatic end-to-end framework for cyberattack construction and emulation. It can autonomously build multi-stage cyberattack plans based on CTI reports, construct the emulation infrastructure, and execute the attack procedures.
\cite{usman2024generativeaitacticalcyber} introduces Occupy AI, a customized and fine-tuned LLM specifically designed to automate and execute cyberattacks. This specialized AI-driven tool is proficient in crafting attack steps and generating executable code for various cyber threats, including phishing, malware injection, and system exploitation.

\textbf{LLM-Assisted Phishing Website/Email Generation.}
\cite{10288940} uses LLM to automatically generate advanced phishing attacks. In the proposed attack method, LLMs are used for the following functions: cloning target websites, modifying login forms to capture credentials, obfuscating code, automating domain name registration, and automating script deployment. 
\cite{roy2024chatbots} examines the potential of LLMs like ChatGPT, GPT-4, Claude, and Bard to generate phishing attacks. The study finds that these models can effectively create convincing phishing websites and emails, mimicking well-known brands and employing evasive tactics to avoid detection. The research also develops a BERT-based detection tool that achieves high accuracy in identifying malicious prompts, serving as a countermeasure against the misuse of LLMs for phishing scams.
\cite{francia2024assessingaivshumanauthored} compares the effectiveness of smishing (SMS phishing) messages created by GPT-4 and human authors, demonstrating that LLM-generated messages are generally perceived as more convincing than those authored by humans. The study also finds that targets are unable to identify whether a message was AI-generated or human-authored and struggle to pinpoint criteria that could help make this distinction. This poses a challenge against personalized AI-enabled social engineering attacks.

\textbf{LLM-Assisted Privilege Escalation Attacks.}
\cite{happe2023evaluating} uses LLM to assist in completing penetration tests. They develop an automated Linux privilege escalation benchmark to evaluate the performance of different LLMs. At the same time, they design a tool called Wintermute to quickly explore the ability of LLMs to bootstrap privilege escalation. %

\textbf{LLM-Assisted Payload Generation.}
\cite{charan2023text} proposes to write payloads with the help of LLMs to launch cyber attacks. This study shows the high efficiency of LLMs by generating executable code for the top 10 MITRE weaknesses observed in 2022 using ChatGPT and Bard respectively. In addition, LLM-generated payloads tend to be more complex and targeted than manually crafted payloads.

\textbf{LLM-Assisted Attack Graph Generation.}
\cite{prapty2024usingretrieveraugmentedlarge} explores the approach of leveraging LLMs to automate the generation of attack graphs by intelligently chaining CVEs based on their preconditions and effects. They also show how to utilize LLMs to create attack graphs from threat reports.

\textbf{LLM-Assisted Capture The Flag (CTF) Challenges.}
\cite{tann2023using} investigates the potential of existing LLMs in solving CTF competitions. They select a number of representative challenges from common CTF categories to evaluate the performance of LLMs, including GPT-3.5, PaLM2, and Prometheus. Their research results demonstrate that LLMs can indeed help participants cope with CTF challenges to a certain extent, albeit not comprehensively.

\textbf{Proxies for Attacks.} 
\cite{beckerich2023ratgpt} uses ChatGPT as a proxy between the victim and the network controlled by the attackers (C\&C), which allows the attacker to remotely control the victim's system without communicating directly, making it difficult to track down the attackers. %

\subsection{(In)secure Code Generation}
\label{subsec:(In)secure Code Generation}

There have been many previous works that have confirmed that LLMs do have good code comprehension capabilities~\cite{he2024instruction,luo2023wizardcoder,roziere2023code,li2023starcoder}. However, the security of the generated code is very important, and some studies have explored this issue.

\textbf{Evaluation of the security of LLM-generated code.}
It is very important to know whether the code generated by LLMs has security risks.
\cite{287298} conducts an experiment to explore whether code written by undergraduate computer science students with the help of LLMs poses any additional security risks. Participants are tasked with implementing a singly-linked 'shopping list' structure in C and they are divided into two groups: a control group that doesn't have access to Codex, and an assisted group that does. The results show that LLM does not significantly increase the risk of introducing security vulnerabilities when used as a code assistant. 
\cite{tambon2024bugs} conducts an empirical study investigating bugs in code generated by LLMs, focusing on three models: CodeGen, PanGu-Coder, and Codex. The research identifies 10 unique bug patterns among 333 collected errors, and these patterns are confirmed by 34 LLM practitioners and researchers. %
\cite{tihanyi2024neutralpromptsproduceinsecure} study how LLMs generate vulnerabilities when writing simple C programs using a neutral zero-shot prompt. They collected code generated by Gemini-pro, GPT-4, Falcon-180B, CodeLLama2-13B, and other LLMs under neutral prompts, which constitute the FormAI-v2 dataset. The study found that at least 63.47\% of the generated programs are vulnerable, highlighting the risks of using LLM-generated code.

There are many studies exploring the security of code generated by state-of-the-art LLMs.
\cite{9833571} investigates the security of code generated by GitHub Copilot. They design 89 different execution scenarios for Copilot, resulting in 1,689 programs. These programs are then analyzed for vulnerabilities, particularly focusing on the top 25 CWEs identified by MITRE.
\cite{wang2024aigeneratedcodereallysafe} introduces CodeSecEval, a meticulously curated dataset designed to address 44 critical vulnerability types with 180 distinct samples. The dataset is then used for precisely evaluating and enhancing the security aspects of code generated by LLMs. The study reveals that current models frequently overlook security issues during both code generation and repair processes, leading to the creation of vulnerable code. 
\cite{wang2023effectiveness} delves into the potential of LLMs in security-oriented program analysis. Their evaluation focuses on two representative LLMs, ChatGPT and CodeBERT, evaluating their performance on analysis tasks of varying difficulty, including vulnerability analysis, bug fixing, fuzzing, and assembly code analysis.
\cite{liu2023need} evaluates the code generated by ChatGPT, focusing on aspects such as correctness, understandability, and security. Through an empirical study using LeetCode questions and CWE scenarios, they analyze the quality of code snippets generated by ChatGPT and its ability to improve the code through multi-round dialogue. The results reveal that while ChatGPT is able to generate functionally correct code, it encounters challenges in complex reasoning and ensuring code security.

On the other hand, \cite{siddiq2023generate} proposes a framework called SALLM specifically for evaluating the security of code generated by LLMs. SALLM consists of three components: a prompt dataset detailing Python programs, a code generation environment that requires different solutions from LLMs, and a systematic evaluation model that leverages Docker to execute the generated code.  %
\cite{liu2024your} focuses on enhancing the quality evaluation of code generation. Recognizing that existing benchmarks often have a limited set of test cases, they introduced a code synthesis evaluation framework, EvalPlus. EvalPlus significantly expands the number of test cases in the evaluation dataset by deploying an automatic test input generator that combines LLMs with a mutation-based strategy.
\cite{ullah2023large} collects 228 code scenarios and analyze 8 LLMs in an automated framework to determine whether LLMs can reliably identify security-related vulnerabilities. They point out that current LLMs fall short in automated vulnerability detection tasks and outline several limitations exhibited by current LLMs.
\cite{buscemi2023comparative} evaluates the performance of ChatGPT-3.5 on generating code, including an examination of code security in 10 programming languages.

\textbf{Do LLMs know whether the generated code is safe or not?}
\cite{khoury2023secure} conducts a series of experiments to evaluate the security of LLM-generated code and to discover vulnerabilities in generated code under various scenarios. The results show that while LLMs may identify vulnerabilities in the generated code when prompted for review, they still generate unsafe code unless explicitly instructed otherwise. 
A significant challenge they faced stems from the uninterpretability of deep neural networks, which causes LLMs to give inconsistent responses when repeatedly asked about code security, without a clear strategy to maximize successful identification.

To ensure the generation of secure codes, LLMSecGuard \cite{kavian2024llm} enhance code security through the synergy between static code analyzers and LLMs. 
\cite{He_2023} takes a more direct approach to customize LLMs through specific mechanisms.
They propose a method named svGen, which makes LLMs generate safe or unsafe code based on the user's security preferences. In addition to the descriptions for the generated code, they also introduce property-specific continuous vectors (called prefixes), which are sequences of vectors that match the shape of the LLMs' hidden states. These prefixes are optimized to influence the LLM's generation process by setting initial hidden states that steer the code toward meeting the desired security criteria, all without modifying the underlying weights of the LLM.

Fine-tuning LLMs for secure code generation is feasible. \cite{li2024exploratorystudyfinetuninglarge} reveals that fine-tuning LLMs can improve secure code generation by 6.4\% for C language and 5.4\% for C++ language. Additionally, fine-tuning with function-level and block-level datasets achieves the best performance in secure code generation, compared to file-level and line-level datasets.
\cite{he2024instruction} introduces SafeCoder, an innovative instruction tuning approach that enhances the security of code generation by LLMs. SafeCoder combines traditional instruction tuning with security-specific fine-tuning using a high-quality dataset collected through an automated pipeline from GitHub. This approach significantly enhances code security without compromising the LLMs' utility across various tasks, demonstrating its adaptability and effectiveness in enhancing the security of LLM-generated code.

In addressing the question of how to best iteratively refine code, \cite{tang2024coderepairllmsgives} points out that the process exposes an explore-exploit tradeoff, which can be framed as a multi-armed bandit problem, and solved using Thompson Sampling. The resulting LLM-based program synthesis algorithm is widely applicable.
\cite{wong2024investigatingtransferabilitycoderepair} discusses iterative code repair in both high and low-resource languages, where an LLM fixes an incorrect program by reasoning about errors and generating new code. Specifically, they delve into guiding the model to generate secure code through chain-of-thought reasoning.

\subsection{Others}
\label{subsec:Others}

Apart from the previously described categories, there are a few scattered studies on the application of LLMs in the field of cybersecurity, which are also of research value.

\textbf{IoT Fingerprint.} \cite{10.1145/3618257.3624845} proposes a method for Internet devices fingerprint generation. Their approach is divided into two steps. First, raw text data obtained from web scans is converted into a stable embedded representation with RoBERTa. Next, the embedding is clustered using the HDBSCAN and the fingerprint is generated based on the clustering.

\textbf{Botnet.} \cite{yang2023anatomy} introduces a LLM-driven botnet called fox8 on Twitter. The fox8 botnet contains over one thousand users controlled by AI. They post machine-generated content and stolen images to spread fake and harmful information, engaging with each other through replies and retweets.

\textbf{Security Patch Detection.} \cite{tang2023justintime} proposes a system named LLMDA, whose main goal is to improve the identification of security patches in open-source software (OSS). LLMs are used to generate explanatory descriptions of patches and synthetic data, which helps to augment existing datasets. %

\textbf{SoC Security.} \cite{saha2023llm} explores the potential of integrating LLMs into the system-on-chip (SoC) security verification paradigm. They provide a systematic evaluation of LLM applications about vulnerability insertion, security assessment, security verification, and countermeasure development. %

\textbf{Taint Analysis.} \cite{liu2023harnessing} introduces LATTE, a static binary taint analysis tool supported by LLMs. LLMs help to identify the chain of data dependencies between taint sources and possible vulnerability triggers. LLMs could provide an understanding of code structure and semantics in the process.

\textbf{LLMs' Input-Output Safeguard.} \cite{inan2023llama} proposes Llama Guard to detect the risk in LLM's prompt and response. Using labeled security risk text, they perform instruction tuning on
Llama2-7b to obtain this model. %

\textbf{Honeypot.} \cite{sladić2024llm} designs a dynamic and real-time fake honeypot by giving response generated by LLMs, which mainly focus on changing the limitation that honeypots are easily recognizable. In their experiment, most people can't recognize whether the remote host is a real one or a honeypot generated by LLMs. %
\cite{reti2024acthoneytokengeneratorinvestigation} systematically investigates the use of LLMs to create a variety of honeytokens. They design different types of honeytokens to evaluate the optimal prompts, including configuration files, databases, and log files. They test 210 different prompt structures, based on 16 prompt-building blocks, and demonstrate that LLMs can generate a wide array of honeytokens using the presented prompt structures.
LLMPot~\cite{vasilatos2024llmpotautomatedllmbasedindustrial} is a novel approach for designing honeypots in ICS networks that harnesses the power of LLMs. It aims to automate and optimize the creation of realistic honeypots with vendor-agnostic configurations, applicable to any control logic, thereby eliminating the manual effort and specialized knowledge traditionally required.

\textbf{Incidence Response.} 
\cite{hays2024employing} advocates for the application of ChatGPT to enhance incident response planning (IRP) in cybersecurity. It suggests that LLMs can draft initial plans, recommend best practices, and identify documentation gaps. The paper highlights the potential of LLMs to streamline IRP processes, emphasizing the value of human oversight to ensure accuracy and relevance. 

\textbf{Network Management.} \cite{10.1145/3626111.3628183} explores how LLMs can be used to generate task-specific code from natural language queries to improve network management. They develop and release a test benchmark, NeMoEval, covering two network management applications: network traffic analysis and network lifecycle management.%

\textbf{Vulnerabilities Reproduction.} \cite{feng2023prompting} proposes an approach called AdbGPT that utilizes LLMs to automatically reproduce vulnerabilities in vulnerability reports by prompting engineering without training or hard coding. %

\textbf{Expertise Q\&A on cybersecurity domain.} \cite{kabir2024stack} conducts an empirical study of ChatGPT's performance in answering Stack Overflow programming questions. The main drawbacks of the LLM answers are fake information and excessive length of the content. Still some testers like its comprehensiveness and good style of language presentation. Due to the difficulty of recognizing misleading information given by LLMs, this is an area that has yet to be researched. %

\begin{tcolorbox}[
  boxrule=0.5pt,
]

Answer to Q2: LLMs have shown great potential in the field of cybersecurity, assisting in various aspects such as threat intelligence, anomaly detection, vulnerability detection, and so on. LLM security copilot can effectively empower the automation and intelligence of cybersecurity, helping to address security risk challenges. Although relevant research has made certain progress, it is still worth further exploration to better apply LLMs in the field of cybersecurity.

\end{tcolorbox}
\section{RQ3: What are the challenge and further research for the application of LLMs in cybersecurity?}
\label{sec:q3}

\begin{figure*}[h]
    \centering 
    \includegraphics[width=\textwidth]{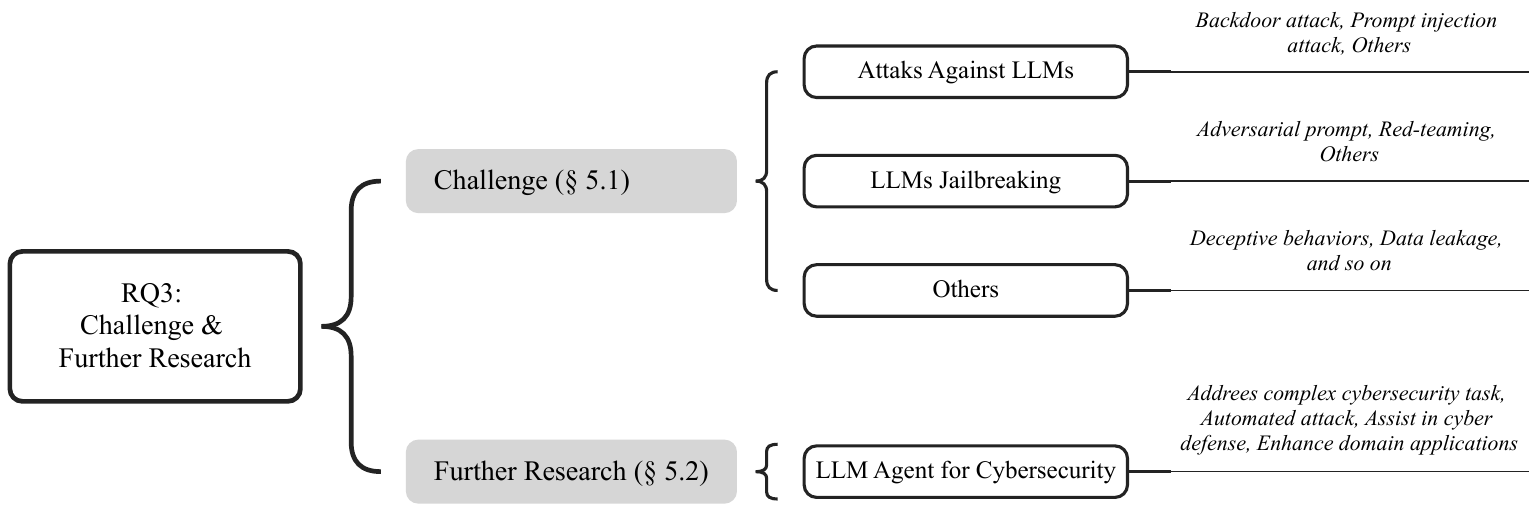}
    \caption{\textbf{An overview of RQ3.}}
    \label{fig:q3}
\end{figure*}

\subsection{Challenge}
\label{subsec:challenge}

The application of LLMs in cybersecurity represents a cutting-edge field, demonstrating the power of LLMs in dealing with complex and dynamic cyber threats. \textbf{However, despite their strengths, LLMs are not without challenges, especially their inherent vulnerabilities and susceptibilities to attacks}~\cite{yao2024survey,zhao2024weaktostrong}. Among the critical concerns are the phenomena of LLMs-oriented attacks and LLMs jailbreaking. These vulnerabilities highlight the double-edged nature of LLM applications in cybersecurity. On one hand, the powerful comprehension and predictive capabilities of LLMs can significantly promote the intelligence of cybersecurity systems. On the other hand, their intrinsic weaknesses facilitate exploitation and pose serious security risks, undermining their reliability and integrity in cybersecurity applications.

We delve into these challenges from two key perspectives: attacks against LLMs, which examines the susceptibility of LLMs to various forms of attacks~\cite{kumarstrengthening,esmradi2023comprehensive,wu2024new}, and LLMs Jailbreaking, focusing on the phenomenon of LLMs generating unsafe or unintended content when prompted in certain ways, despite being designed with safeguards~\cite{chu2024comprehensive,xu2024llm}. Through an analysis of these dimensions, we aim to illuminate the complexities of exploiting LLMs in cybersecurity, highlighting the need for caution and strategic foresight in their application.

\textbf{Attaks Against LLMs.} 
The vulnerabilities of LLMs make them susceptible to attacks by malicious users. We focus on two types of attacks: backdoor attacks and prompt injection attacks.

\textit{Backdoor Attack} manipulates model outputs to achieve attackers' objectives by embedding specific triggers in the model or its inputs. 
\cite{shi2023badgpt} proposes a novel backdoor attack methodology called BadGPT, specifically targeting language models that have been fine-tuned through reinforcement learning, such as ChatGPT. This approach involves embedding backdoors within the reward model, which can be activated via specific trigger prompts. Such activation allows attackers to control the model's output to align with their preferences, showcasing a critical security vulnerability. 
In another study, \cite{zhao2024universal} introduces a novel backdoor attack strategy, ICLAttack, which aims at exploiting the inherent context learning capabilities of LLMs. The ICLAttack framework encompasses two primary attack vectors: poisoning demonstration examples and poisoning demonstration prompts. By embedding backdoor triggers within the model's context, ICLAttack is able to influence the model's behavior without the need for fine-tuning, thus revealing universal vulnerabilities within LLMs. 
Furthermore, \cite{yao2023poisonprompt} reveals a backdoor attack mechanism tailored to prompt-based LLMs, called PoisonPrompt. The method injects backdoors into the language model through two steps: poisoned prompt generation and bi-level optimization. PoisonPrompt can alter the normal prediction of the model in case of specific trigger activations without affecting the performance of the model on downstream tasks, posing a subtle but powerful threat to the integrity of LLMs.

\textit{Prompt Injections Attack} involve attackers inserting malicious commands into inputs, compelling the model to execute actions aligned with the attackers' intentions.
\cite{pedro2023prompt} conducts a comprehensive investigation of prompt-to-SQL (P2SQL) injection attacks against web applications based on the Langchain framework. These attacks utilize user-input prompts to generate malicious SQL queries, thereby enabling attackers to tamper with databases or steal sensitive information. 
\cite{jiang2023prompt} introduces the Compositional Instruction Attack (CIA), unveiling the susceptibility of LLMs to attacks that utilize synthetic instructions with potentially malicious intentions. Through two transformation methods, Talking-CIA and Writing-CIA, harmful instructions are masked as conversational or writing tasks, preventing the model from recognizing potentially malicious intent and thus generating harmful content. 
\cite{liu2023prompt} proposes a novel black-box prompt injection attack technique named HOUYI for applications integrated with LLMs. HOUYI executes attacks through three key elements: pre-constructed prompts, injection prompts, and malicious payloads. Its deployment across 36 real-world scenarios demonstrates its efficacy in discovering and exploiting vulnerabilities within LLM-integrated applications.  
\cite{yan2023backdooring} focuses on Virtual Prompt Injection (VPI) attacks against instruction-tuned LLMs, which allow attackers to manipulate model behavior by specifying virtual prompts without directly injecting into model inputs, leading to the model disseminating biased information. 
\cite{piet2024jatmo} uses instruction-tuned models to generate datasets for specific tasks. These datasets are then utilized to fine-tune foundational models, enhancing their robustness to resist most prompt injection attacks.

Additionally, \cite{kour2023unveiling} constructs an adversarial attack dataset named AttaQ in a semi-automated manner, aiming to evaluate the security of LLMs in the face of harmful or inappropriate inputs. Vulnerabilities are exposed by analyzing model responses to the AttaQ dataset, and specialized clustering techniques are further applied to identify and characterize the models' vulnerable semantic areas.
\cite{esmradi2023comprehensive} conducts a comprehensive survey of various attack types targeting LLMs, encompassing both direct attacks on the models themselves and indirect attacks on applications utilizing the models. This study describes the impacts of these attacks on the privacy, security, and reliability of the models. And it underscores the critical importance of implementing proactive security measures in the development of AI models.

\textbf{LLMs Jailbreaking.} 
As mentioned above, LLM is susceptible to various attacks, with jailbreaking attacks being one of the most popular. 
\cite{shen2023do} studies the security issues of LLMs when facing jailbreak prompts. They collect and analyze 6,387 prompts to reveal the characteristics and attack strategies of these prompts. Despite various security measures implemented by LLMs, they found that effective jailbreak prompts still successfully induce models to generate harmful content, indicating the need for further improvements in the security of LLMs. 
\cite{chu2024comprehensive} conducts a comprehensive evaluation of LLMs jailbreaking, revealing the effectiveness of these attack methods and the vulnerabilities of LLMs across various violation categories.

There are various methods for generating adversarial prompts. 
\cite{zou2023universal} combines greedy search and gradient-based optimization techniques to propose a method that automatically generates adversarial suffixes to prompt models, both open-source and commercial, to produce inappropriate content. 
\cite{lapid2023open} introduces a novel approach to black-box jailbreak attacks using genetic algorithms, which can manipulate LLMs to produce unexpected and potentially harmful outputs without accessing the model's internal structure and parameters by optimizing a universal adversarial prompt. 
\cite{ding2024wolf} conceptualizes the jailbreaking process as prompt rewriting and scenario nesting. They then introduce ReNeLLM, a jailbreaking prompt generation framework that utilizes LLMs to generate effective jailbreaking prompts. Compared to existing baselines, ReNeLLM achieves high attack success rates on multiple LLMs while significantly reducing the time cost.
\cite{Deng_2024} explores jailbreak attacks on LLM Chatbots and proposes a framework named MASTERKEY to automate this process. Through temporal feature analysis and automated prompt generation, MASTERKEY reveals and bypasses the defense mechanisms of LLM chatbots, offering new perspectives for LLM security research and guidance for service providers to improve their security measures.

Research on LLMs jailbreaking can also be used for red-teaming. 
\cite{zhu2023autodan} proposes AutoDAN, an interpretable and gradient-based adversarial attack method. By combining the dual objectives of jailbreaking and readability, it generates interpretable and diverse attack prompts capable of effectively bypass perplexity filters and demonstrates robust generalization in scenarios with limited training data. This method not only offers a novel approach for red-teaming of LLMs but also helps to understand the mechanics of jailbreak attacks. 
\cite{yu2023gptfuzzer} presents a new black-box jailbreak fuzzing framework named GPTFUZZER. By collecting human-written jailbreak templates from the internet as initial seeds, and then iterating through a process of seed selection, mutation, and evaluating the success of attacks, GPTFUZZER significantly enhances the efficiency and scalability of red team testing. 
\cite{yao2023fuzzllm} introduces FuzzLLM, a novel and universally applicable fuzz testing framework designed to proactively discover jailbreak vulnerabilities in LLMs. FuzzLLM employs a template-based strategy that generates a variety of jailbreak prompts and identify potential security vulnerabilities through automated testing. It demonstrates efficiency and comprehensiveness across various LLMs, effectively identifying and assessing jailbreak vulnerabilities.

Additionally, \cite{wang2023selfdeception} introduces the concept of a semantic firewall to describe the defense mechanisms of LLMs against malicious prompts and proposes a self-deception attack method to bypass LLMs semantic firewalls. This method designs a customizable dialogue template for experimenting with specific illegal payloads and automatically achieving LLM jailbreak. 
\cite{qiu2023latent} develops a potential jailbreak prompt dataset embedded with malicious instructions and proposes a hierarchical annotation framework to analyze the performance of LLMs under different conditions(\emph{e.g.}, instruction positions, word substitutions, and instruction replacements). This is aimed at evaluating the security and output robustness of LLMs when processing texts containing potential malicious instructions. 
\cite{li2023multistep} investigates the potential privacy threats associated with ChatGPT and the Bing search engine integrated with ChatGPT. By introducing a novel multi-step jailbreaking prompt, they successfully extract personally identifiable information from ChatGPT and demonstrate the privacy threats posed by the new Bing under direct prompts.  

\textbf{Others.} 
Besides the extensively researched vulnerabilities, several other LLM risks limit their application in cybersecurity. 
\cite{10375472} highlights the dual-edged nature of generative AI and ChatGPT, revealing that while they bring convenience, they also pose cybersecurity and ethical challenges. 
\cite{hubinger2024sleeper} investigates the deceptive behaviors that LLMs may exhibit under specific trigger conditions and finds that these behaviors might persist even after safety alignment, posing a potential threat to the security of AI systems. 
\cite{yang2023shadow} points out that even securely aligned LLMs can be easily manipulated to generate harmful content with simple data tuning, highlighting the complexity of maintaining LLM security. 
\cite{jiang2023identifying} identifies critical vulnerabilities within LLM-integrated applications, which could stem from malicious app developers or external threats with the capability to control database access, manipulate, and polluting data. 
\cite{sallou2023breaking} also raises data leakage and reproducibility issues associated with the use of closed-source LLMs.

\begin{tcolorbox}[
  boxrule=0.5pt,
]

Answer 1 to Q3: Despite the powerful capabilities of LLMs, they inherently possess certain weaknesses and vulnerabilities, making them susceptible to attacks. In particular, jailbreaking poses significant security risks to the application of LLMs.

\end{tcolorbox}

\subsection{Further Research}
\label{subsubsec:further_research}

Despite the significant research into LLMs within the field of cybersecurity, the exploration and application of such models remain in their initial stages and have great potential for development~\cite{dasilva2024survey,motlagh2024large}. The complexity of cybersecurity stems not only from the diversity of attack methods but also from the intricate nature of network environments, which requires the integrated application of various tools and strategies to achieve effective protection~\cite{azizi2023cybersecurity,mtsweni2018unified}. Facing these challenges requires AI systems to have stronger capabilities in planning, reasoning, tool use, and memory. \textbf{Consequently, the concept of LLM Agent has emerged and attracted a lot of attention from researchers.}

LLM Agent is ``a system that can use an LLM to reason through a problem, create a plan to solve the problem, and execute the plan with the help of a set of tools~\cite{nvidia23agent}.'' By simulating complex network behaviors and attack patterns, and integrating advanced natural language processing capabilities, LLM agents introduce new perspectives and solutions to the field of cybersecurity~\cite{kaheh2023cyber,moskal2023llms,cui2024llmind,rigaki2023cage,fang2024llm,an2024nissist}. With the continuous advancement of technology and in-depth research, LLM agents are expected to play a key role in defense strategy generation, threat detection, and security policy formulation, significantly improving the efficiency and intelligence level of cybersecurity defenses.

The AI Agents framework based on LLMs possesses the critical capabilities required to solve complex problems~\cite{ruan2023tptu}. 
\cite{xi2023rise} proposes an LLM Agent architecture that includes brain, perception, and action components to provide a wide range of applications in single-agent scenarios, multi-agent environments, and human-agent collaboration. 
Moreover, the incorporation of Tool \& API calls endows LLM agents with the capacity to interact with the real world. 
\cite{qin2023toolllm} develops the ToolBench dataset and the DFSDT algorithm to enable LLMs to successfully handle complex tasks involving numerous real-world APIs. 
\cite{liu2024summary} introduces a sophisticated tool invocation mechanism that enhances LLMs' interaction with external tools by summarizing and making decisions.
Additionally, \cite{yang2024llm} demonstrates that integrating code into LLMs significantly enhances its ability to perform more complex tasks as an intelligent agent.
\cite{qiao2023taskweaver} proposes TaskWeaver, a code-first agent framework for seamlessly planning and executing data analytics tasks.

LLM agents can be applied to address complex cybersecurity tasks.
\cite{cui2024llmind} proposes an innovative framework named LLMind, which utilizes LLM as a coordinator to perform complex tasks by integrating with IoT devices and domain-specific AI modules. The framework employs finite state machine methods to generate control scripts, thereby enhancing the accuracy and success rate of task execution. In addition, LLMind introduces a mechanism for accumulating experience, which allows the system to continually learn and progress through ongoing interactions between users and machines. 
\cite{rigaki2023cage} demonstrates the use of LLMs as agents within cybersecurity environments. Experiments show that LLM agents can achieve performance comparable to or better than extensively trained agents in sequential decision-making tasks, even without additional training. Furthermore, the study introduces the NetSecGame environment, a highly modular and adaptive cybersecurity environment designed to support complex multi-agent scenarios.
\cite{huang2023large} proposes ChatNet, a domain-specific network LLM framework with access to a variety of external network tools. ChatNet significantly reduces the time required for tedious network planning tasks, thereby greatly increasing efficiency.

LLM agents can be employed to perform automated attacks. 
\cite{fang2024llm} reveals the potential of LLM agents in cybersecurity attacks, particularly the capability of GPT-4 to autonomously conduct complex hacker attacks on websites without prior knowledge of vulnerabilities. The study shows that LLM agents have a success rate of up to 73.3\% in hacking attempts and can autonomously discover vulnerabilities in real-world websites. 
\cite{moskal2023llms} demonstrates the potential application of LLMs in cyber threat testing, especially in automating cyber attack activities. With prompt engineering and automated agents, LLMs can understand and execute complex cyber attacks. 
\cite{fang2024llmagentsautonomouslyexploit} collects a dataset of 15 zero-day vulnerabilities. Based on this dataset, the study shows that LLM agents can autonomously exploit these zero-day vulnerabilities in real-world systems.
\cite{fang2024teamsllmagentsexploit} also shows that teams of LLM agents can exploit real-world, zero-day vulnerabilities by designing a system of agents with a planning agent that can launch subagents.

LLM agents can also be utilized to assist in cyber defense. 
\cite{an2024nissist} designs a multi-agent system (Nissist) to precisely understand user queries and provide effective mitigation plans. Nissist utilizes troubleshooting guides and incident mitigation history to provide suggestions, which significantly reduces the time for event mitigation, reduces the workload of on-duty engineers, and enhances the reliability of services. 
Cyber Sentinel~\cite{kaheh2023cyber} is a dialogue agent based on GPT-4, which can interpret potential cyber threats and execute security actions based on user instructions. The potential impact of Cyber Sentinel in cyber security includes improved threat detection and response capabilities, enhanced operational efficiency, real-time collaboration, and knowledge sharing.
PhishAgent~\cite{cao2024phishagentrobustmultimodalagent} is a multimodal agent that combines a wide range of tools, integrating both online and offline knowledge bases with Multimodal LLMs, showing strong resilience against various types of adversarial attacks.
\cite{tseng2024usingllmsautomatethreat} develops an AI agent to replace the labor intensive repetitive tasks involved in analyzing CTI reports. By leveraging the advanced capabilities of LLMs, the AI agent can accurately extract important information from large volumes of text and generate Regex to help SOC analysts accelerate the process of establishing correlation rules.

LLM agents enhance cybersecurity applications with their remarkable capabilities, yet the security risks inherent in agent systems~\cite{yuan2024rjudge} pose challenges for their deployment in cybersecurity environments. 
\cite{wu2024wipi} introduces the concept of Web-based Indirect Prompt Injection (WIPI), a novel cyber threat that embeds malicious instructions in web pages to indirectly control these agents, achieving high success rates and robustness across different user inputs. 
\cite{zhan2024injecagent} highlights that LLM agents integration with external tools may lead to the risk of indirect prompt injection attacks, in which attackers embed malicious commands in the content processed by LLMs to manipulate these agents to perform actions harmful to users.

In conclusion, the application of LLM-based agents in cybersecurity opens up new avenues for dealing with cyber security threats. Although research in this area is still in its early stages, and the inherent security vulnerabilities of agents have not yet been addressed, this line of research promises to significantly enhance the capability to counter complex cyber threats and has the potential to revolutionize the working methods of security professionals, thereby unleashing greater productivity. Therefore, further research into the application of LLM agents in cybersecurity is crucial for developing adaptive, intelligent, and comprehensive cybersecurity solutions.

\begin{tcolorbox}[
  boxrule=0.5pt,
]

Answer 2 to Q3: Extending the tool-use and API-call capabilities of LLM, coupled with the design of autonomous intelligent agents capable of understanding, planning, and executing complex tasks within cybersecurity applications, will greatly advance the utilization of AI in cybersecurity.

\end{tcolorbox}

\section{Conclusion}
\label{sec:conclusion}

This paper introduces the methodologies for constructing cybersecurity-oriented domain LLMs, detailing how existing models can be fine-tuned to meet specific needs using target data. 
The investigation into the applications of LLMs has shows that LLMs have great potential for a wide range of cybersecurity tasks, such as threat intelligence, vulnerability detection, secure code generation and others. 
However, we has also acknowledged the inherent vulnerabilities of LLMs, particularly the susceptibility to attacks such as jailbreaking, which pose significant security risks. Mitigating these vulnerabilities is crucial to securely deploying LLMs in sensitive environments. 
Additionally, we propose future research directions, such as extending the tool-use and API-call capabilities of LLMs, and developing autonomous intelligent agents for complex cybersecurity operations.

In summary, we bridges the gap between LLM advancements and cybersecurity demands, laying the groundwork for researchers and practitioners. It guides them to harness the transformative potential of LLMs while addressing the unique challenges that arise in this field. 
Further research and exploration would open up new pathways for future cybersecurity practice, ensuring that we have more comprehensive and professional strategies in the face of increasingly complex cyber threats.

\clearpage
\bibliographystyle{unsrt}  
\bibliography{references}

\end{document}